\documentclass[aps,pre,epsf,superscriptaddress,amsmath,amssymb,amsfonts,twocolumn,showpacs]{revtex4-1}

\usepackage{graphicx}
\usepackage{epsfig}
\usepackage{dcolumn}
\usepackage{bm}
\usepackage{braket}
\usepackage{amsmath}
\usepackage{color}

\begin{document}
\title{Correlation Effects in the Quench-Induced Phase Separation Dynamics\\ of 
a Two-Species Ultracold Quantum Gas}

\author{S. I. Mistakidis}
\affiliation{Zentrum f\"{u}r Optische Quantentechnologien,
Universit\"{a}t Hamburg, Luruper Chaussee 149, 22761 Hamburg,
Germany} 
\author{G. C. Katsimiga}
\affiliation{Zentrum f\"{u}r Optische Quantentechnologien,
Universit\"{a}t Hamburg, Luruper Chaussee 149, 22761 Hamburg,
Germany}
\author{P. G. Kevrekidis}
\affiliation{Department of Mathematics and Statistics, University
of Massachusetts Amherst, Amherst, MA 01003-4515, USA }
\author{P. Schmelcher}
\affiliation{Zentrum f\"{u}r Optische Quantentechnologien,
Universit\"{a}t Hamburg, Luruper Chaussee 149, 22761 Hamburg,
Germany} \affiliation{The Hamburg Centre for Ultrafast Imaging,
Universit\"{a}t Hamburg, Luruper Chaussee 149, 22761 Hamburg,
Germany}

\date{\today}

\begin{abstract}
 
We explore the quench dynamics of a binary Bose-Einstein 
condensate crossing the miscibility-immiscibility threshold and vice versa,
both within and in particular beyond the mean-field approximation.
Increasing the interspecies repulsion leads to the filamentation of the 
density of each species, involving shorter   
wavenumbers and longer spatial scales in the many-body approach.
These filaments appear to be strongly correlated  
and exhibit domain-wall structures.
Following the reverse quench process multiple dark-antidark solitary waves  
are spontaneously generated and subsequently
found to decay in the many-body
scenario. We simulate single-shot images to connect our findings to possible experimental
realizations. Finally, the growth rate of the variance of a sample of single-shots probes the degree 
of entanglement inherent in the system. 
\end{abstract}

%\pacs{Pacs numbers} 
\maketitle

\section{Introduction}

The realm of atomic Bose-Einstein condensates (BECs) has offered over
the past two decades a fertile testbed for the examination of
phenomena involving the role of nonlinearity in wave dynamics
and phase transitions~\cite{pethick,stringari,emergent,darkbook,lcarr,prouka}.
Phase separation dynamics in the case of multi-component BECs
has held a prominent role among the relevant studies and is
a topic that by now has been summarized in various reviews~\cite{pethick,stringari,darkbook,ueda}.
Nevertheless, the majority of the relevant studies has focused on
a mean-field (MF) description, while the role
of many-body (MB) effects in such transitions is much less
understood.

Since the early days of the experimental realization of BECs,
experimental achievements include binary mixtures of e.g. two
hyperfine states of $^{23}$Na~\cite{nake}  and of~$^{87}$Rb~\cite{dsh}. 
Progress of the experimental control over the relevant
multi-component settings enabled detailed observations of phase separation phenomena and related dynamical 
manifestations~\cite{stenger,cornell,usadsh,wieman,hall2,hall3,tojo,eto1,eto2}. 
In recent years, external coupling fields have been utilized
to control and modify the thresholds for mixing-demixing dynamics in pseudo-spinor (two-component)~\cite{flop,strobel} and 
even in spinor systems~\cite{spielman}. Moreover, the quench dynamics across
the phase separation transition has been a focal point of
studies examining the scaling properties of suitable correlation
functions and associated universality properties~\cite{gas1,gas2,bisset}.

More recently the inclusion of correlations in multi-component few boson
systems enabled a microscopic characterization of their static properties. 
A variety of novel phases have been realized in these settings such as altered phase separation 
processes~\cite{March,Cazalilla,Alon_mixt,March1,Mishra}, 
composite fermionization~\cite{comp_ferm,Pyzh,Hao}, or even the crossover between the two~\cite{March2,March3}. 
Also the dynamical properties of such MB ultracold mixtures have been studied including, among others,
the dependence of the tunnelling dynamics on the mass ratio~\cite{Pflanzer,Pflanzer1} or the intra- and interspecies 
interactions~\cite{Chatterjee}, 
as well as the emergence of Anderson's orthogonality catastrophe upon quenching the interspecies repulsion~\cite{Campbell}. 
On the other hand, far less emphasis has been placed on the
MB character of the quench-induced phase separation phenomenology.
It is the latter apparent gap in the literature that the present work aims at addressing
for both few and larger bosonic ensembles. 

To incorporate the quantum fluctuations due to correlations \cite{Mistakidis,Mistakidis1,Mistakidis2,Mistakidis5} emerging when quenching the binary
BEC system,
we bring to bear the Multi-Layer Multi-Configuration 
Time-Dependent Hartree Method for bosons (ML-MCTDHB)~\cite{Lushuai,MLX} 
designed for simulating the quantum dynamics
of bosonic mixtures. We explore different scenarios, emphasizing
the case where the inter-species interaction is quenched
from the miscible to the immiscible regime (positive quench) or vice versa
(negative quench).
We find significant variations in the MB scenario in comparison 
to the MF one. 
In the positive quench scenario the unstable dynamics leads to the filamentation of the density 
of each species and the dominant wavenumber associated
with the emerging phase separated state appears to generically be higher
in the MF case.
The one- and the two-body correlation functions indicate the presence of correlations between the filaments
of the same or different species signalling the presence of fragmentation and entanglement respectively. 
In particular, strong one-body correlations appear between 
non-parity symmetric (with respect to the trap center)
filaments formed indicating their tendency of localization. 
These filaments are found to be strongly 
anti-correlated at the two-body level indicating a negligible probability of finding two bosons of the same species 
one residing in an outer and one in an inner filament. 
More importantly, combining the behavior of one- and two-body correlations supports the formation of domain-walls 
i.e. interfaces that separate these distinct
filaments~\cite{trippenbach,boris,de}.   

In sharp contrast to the above dynamical manifestation of the phase separation,
in the negative quench scenario multiple dark-antidark (DAD), 
i.e. density humps on top of the BEC background, solitary waves~\cite{Danaila,KevrekidisDAD} 
are spontaneously generated both within and beyond the MF approximation. At the MB 
level many decay events, at the early stages of the dynamics, increase the production
of DAD solitary waves with the product of each decay being a slow and a fast DAD structure~\cite{lgs}.
The latter increase results in multiple collisions and interference events between these matter waves,
and most of them are lost during evolution.
Furthermore, in both the positive and the negative quench scenarios,
single-shot simulations, utilized here for the first time for binary mixtures,
offer a link to potential experimental realizations of the above-observed dynamics.
In particular, the growth rate of the variance of single-shots resembles the growth rate 
of the entanglement inherent in the system. 
Additionally, deviations between the variances of 
the two species reveal the fragmented nature of the binary system.  
Last, but not least the case of quenches
within the immiscible regime, are explored showcasing  
the one-dimensional (1D) analogue of the so-called ``ball'' and ``shell'' structure
appearing in higher-dimensional binary BECs~\cite{usadsh}. 
 
Our presentation is structured as follows. In section \ref{setup}, we provide
the details of the binary setup and the corresponding MB
ansatz, briefly addressing the ML-MCTDHB approach. 
In section \ref{quench} we examine the different quench scenarios focusing
on the miscible to immiscible quench as well as the reverse quench dynamics. 
Section \ref{conclusions} provides a summary of our findings and a number
of proposed directions for future study. 
In Appendix A we present the details of the single-shot procedure, and in 
Appendix B we show how the quench induced phase separation dynamics is altered for 
small particle numbers.  
Finally, in Appendix C we address the convergence of the ML-MCTDHB results.

\section{Setup and many-body ansatz}\label{setup}
To explore the correlated out-of-equilibrium quantum dynamics in a relevant experimental setting, 
we consider a binary bosonic gas trapped in a 1D harmonic oscillator potential. 
The MB Hamiltonian consisting of $N_A$, $N_B$ bosons with masses $m_A$, $m_B$ for the species $A$, $B$ respectively,  
reads 
\begin{equation}
\begin{split}
H= &\sum_{\sigma=A,B} \sum_{i=1}^{N_{\sigma}} \left[ -\frac{\hbar^2}{2 m_{\sigma}} \left( \frac{d}{d x^{\sigma}_i} \right)^2 %
+\frac{1}{2} m_{\sigma} \omega_{\sigma}^2  \left( x^{\sigma}_i \right)^2 \right]\\ 
&+\sum_{\sigma=A,B}g_{\sigma\sigma} \sum_{i<j} \delta( x^{\sigma}_i - x^{\sigma}_j)\\ 
&+g_{AB} \sum_{i=1}^{N_A} \sum_{j=1}^{N_B} \delta( x^A_i - x^B_j).
\end{split}
\label{Eq:1}
\end{equation}
In the $s$-wave scattering limit~\cite{pethick} 
both the intra and interspecies interactions are modeled
by a contact potential, where the effective coupling constants are denoted by 
$g_{AA}$, $g_{BB}$, and $g_{AB}$ respectively.
Experimentally $g_{\sigma\sigma'}$ can be tuned 
either via the three-dimensional scattering length with the aid of  
Feshbach resonances~\cite{Kohler,Chin} or via the corresponding transversal 
confinement frequency and the resulting confinement-induced resonances~\cite{Olshanii,Kim}. 
Moreover, here we assume that both species possess the same mass, i.e. $m_A$=$m_B$=$m$,
and are confined in the same external potential, i.e. $\omega_A$=$\omega_B$=$\Omega$. 
Throughout this work the trapping frequency is fixed to $\Omega=0.1\approx 2 \pi \times 20$Hz assuming 
a transversal confinement $\omega_{\perp}=2 \pi \times 200$Hz.
Furthermore, we fix the intraspecies interactions to $g_{AA}=1.004$ and $g_{BB}=0.9544$, which are 
the values for a binary BEC of $^{87}$Rb atoms prepared in the internal states 
$\Ket{F=1, m_F=-1}$ and $\Ket{F=2, m_F=1}$~\cite{hall3}, 
while $g_{AB}$ is left to arbitrarily vary upon a quench taking values within the interval
$g_{AB}=[0,2]$.
We remark that in the following the Hamiltonian of Eq.~(\ref{Eq:1}) is rescaled in harmonic 
oscillator units, $\tilde{H}=H/(\hbar\Omega)$. 
Then the corresponding length, energy, time, and interaction strength are given in 
units of $\sqrt{\hbar/(m\Omega)}$, $\hbar\Omega$, $\Omega^{-1}$, and $g'_{\sigma \sigma'}=g_{\sigma 
\sigma'}\sqrt{m/\hbar^3\Omega}$, respectively.

Within the MF approximation all particle correlations are neglected.
Such a simplification allows for expressing the MB wavefunction of a binary system 
as a product state of the respective MF wavefunctions
\begin{equation}
\begin{split}
\Psi_{MF} (\vec x^A,\vec x^B;t) &= \Psi^A_{MF} (\vec x^A;t) \Psi^B_{MF} (\vec x^B;t)\\
&= \prod_{i=1}^{N_A} \frac{\phi^A(x^A_i;t)}{\sqrt{N_A}} \prod_{i=1}^{N_B} \frac{\phi^B(x_i;t)}{{\sqrt{N_B}}}, \label{Eq:2}
\end{split}
\end{equation}
where $\vec x^{\sigma}=\left( x^{\sigma}_1, \dots, x^{\sigma}_{N_{\sigma}} \right)$ denote the spatial $\sigma=A,B$ 
species coordinates, $N_{\sigma}$ is the number of $\sigma$ species atoms and 
$\phi^{\sigma}(x^{\sigma}_i;t)$ refers to the time-evolved wavefunction for the $\sigma$ species 
within the MF approximation.
Employing a variational principle, e.g. the Dirac-Frenkel one~\cite{Dirac,Frenkel}, 
for the ansatz of Eq.~(\ref{Eq:2}) we obtain the corresponding equations of motion in the form of
the well-studied system of coupled Gross-Pitaevskii equations \cite{pethick,stringari}.

The binary BEC is a bipartite composite system residing in the Hilbert space 
$\mathcal{H}^{AB}=\mathcal{H}^A\otimes\mathcal{H}^B$, 
with $\mathcal{H}^{\sigma}$ being the Hilbert space of the $\sigma$ species.  
To incorporate correlations between the different (inter-) or the same (intra-) species, $M$ distinct 
species functions for each species are introduced obeying $M\le \min(\dim(\mathcal{H}^A),\dim(\mathcal{H}^B))$.  
Then the MB wavefunction $\Psi_{MB}$ can be expressed according to the truncated Schmidt decomposition \cite{Horodecki} of 
rank $M$  
\begin{equation}
\Psi_{MB}(\vec x^A,\vec x^B;t) = \sum_{k=1}^M \sqrt{ \lambda_k(t) }~ \Psi^A_k (\vec x^A;t) \Psi^B_k (\vec x^B;t).  
\label{Eq:3}
\end{equation}
The Schmidt weights $\lambda_k(t)$ in decreasing order are referred to as the natural species populations of the $k$-th 
species function $\Psi^{\sigma}_k$ of the $\sigma$ species. 
We remark that $\{\Psi_k^{\sigma}\}$ forms an orthonormal $N_{\sigma}$-body wavefunction set in a subspace of 
$\mathcal{H}^{\sigma}$.  
To quantify the presence of interspecies correlations or entanglement we use the eigenvalues
$\lambda_k$ of the species reduced density matrix
$\rho^{N_{\sigma}} (\vec{x}^{\sigma}, \vec{x}'^{\sigma};t)=\int d^{N_{\sigma'}} x^{\sigma'} \Psi^*_{MB}(\vec{x}^{\sigma}, 
\vec{x}^{\sigma'};t) \Psi_{MB}(\vec{x}'^{\sigma},\vec{x}^{\sigma'};t)$, where $\vec{x}^{\sigma}=(x^{\sigma}_1), \cdots, 
x^{\sigma}_{N_{\sigma}-1})$, and $\sigma\neq \sigma'$. When only one (multiple) eigenvalue(s) of $\rho^{N_{\sigma}}$ 
is (are) macroscopic the system is referred to as non-entangled (species entangled or interspecies correlated). 
It is also evident from Eq. (\ref{Eq:3}) that the system is entangled \cite{note_ent,Roncaglia} when at least two distinct 
$\lambda_k(t)$ are finite, further implying that the MB state cannot be expressed as a direct product of two states stemming 
from $\mathcal{H}^A$ and $\mathcal{H}^B$. 
In this manner, $1-\lambda_1(t)$ offers a measure for the degree of the system's entanglement. 
Moreover, a particular configuration of $A$ species $\Psi_k(\vec x^A;t)$ is accompanied by a 
particular configuration of $B$ species $\Psi_k(\vec x^B;t)$ and vice versa.  
Indeed, measuring one of the species states e.g. $\Psi_{k'}^{A}$ collapses the wavefunction of the other species 
to $\Psi_{k'}^{B}$ thus manifesting the bipartite entanglement \cite{Peres,Lewenstein1}.   
Concluding, the above MB wavefunction ansatz $\Psi_{MB}$ constitutes an expansion in terms of different interspecies 
modes of entanglement, where $\sqrt{\lambda_k(t)} \Psi^A_k (\vec x^A;t)\Psi^B_k (\vec x^B;t)$ corresponds to the $k$-th 
entanglement mode.  

To include interparticle correlations we further expand each of
the species functions $\Psi^{\sigma}_k (\vec x^{\sigma};t)$ using the permanents of $m^{\sigma}$ distinct 
time-dependent single particle functions (SPFs) namely $\varphi_1,\dots,\varphi_{m^{\sigma}}$    
\begin{equation}
\begin{split}
&\Psi_k^{\sigma}(\vec x^{\sigma};t) = \sum_{\substack{n_1,\dots,n_{m^{\sigma}} \\ \sum n_i=N}} c_{k,(n_1,
\dots,n_{m^{\sigma}})}(t)\times \\ &\sum_{i=1}^{N_{\sigma}!} \mathcal{P}_i
 \left[ \prod_{j=1}^{n_1} \varphi_1(x_j;t) \cdots \prod_{j=1}^{n_{m^{\sigma}}} \varphi_{m^{\sigma}}(x_j;t) \right].  
 \label{Eq:4}
 \end{split}
\end{equation} 
Here, $c_{k,(n_1,\dots,n_{m^{\sigma}})}(t)$ are the time-dependent
expansion coefficients of a particular permanent, $\mathcal{P}$ is the 
permutation operator exchanging the particle configuration within the SPFs, 
and $n_i(t)$ denotes the occupation number of the SPF $\varphi_i(\vec{x};t)$.  
Following the Dirac Frenkel \cite{Frenkel,Dirac} variational principle 
for the generalized ansatz [see Eqs.~(\ref{Eq:3}), (\ref{Eq:4})] yields the ML-MCTDHB equations of motion 
\cite{Lushuai,MLX,note1}. 
These consist of a set of $M^2$ ordinary (linear) differential equations of motion for the coefficients $\lambda_k(t)$,   
coupled to a set of $M(\frac{(N_A+m^A-1)!}{N_A!(m^A-1)!}+\frac{(N_B+m^B-1)!}{N_B!(m^B-1)!})$     
non-linear integrodifferential equations for the species functions, and $m_A+m_B$ nonlinear 
integrodifferential equations for the SPFs.   

According to the above MB expansion, the one-body reduced density matrix of $\sigma$ species can be expanded in 
different modes [see Eq.~(\ref{Eq:3})]  
\begin{equation}
\begin{split}
&\rho^{(1),{\sigma}}(x,x';t)=\int d^{N_{\sigma}-1}\bar x^{\sigma} d^{N_{\sigma'}} x^{\sigma'} \times ~\\&\Psi^{*}_{MB}(x,\vec{\bar x}^{\sigma},\vec x^{\sigma'};t) 
\Psi_{MB}(x',\vec{\bar x}^{\sigma},\vec x^{\sigma'};t)\\
&=\sum_{k=1}^M \lambda_k(t)~ \rho^{(1),\sigma}_k(x,x';t),
\end{split} \label{Eq:5}
\end{equation} 
where $\sigma \neq \sigma'$, $\bar x^{\sigma}=(x^{\sigma}_1,x^{\sigma}_2,\ldots,x^{\sigma}_{N_{\sigma}-1})$, 
and $\rho^{(1),\sigma}_i(x,x';t)=\int d^{N_{\sigma} -1} \bar{x}^{\sigma} \Psi^{* \sigma}_i (x,\bar{x}^\sigma;t) 
\Psi^{\sigma}_i (x',\bar{x}^\sigma;t)$ 
denotes the one-body density matrix of the $i$-th species function. 
Note here that the system is termed intraspecies correlated or fragmented if multiple 
eigenvalues of $\rho^{(1),\sigma}(x,x')$ are macroscopically occupied, 
otherwise is said to be fully coherent or condensed.

The eigenfunctions of the one-body density matrix $\rho^{(1),{\sigma}}(x,x')$ are the 
so-called natural orbitals $\phi^{\sigma}_i(x;t)$. 
Here we consider them to be normalized to their corresponding eigenvalues, $n^{\sigma}_i$ (natural populations)  
\begin{equation}
n^{\sigma}_i(t)= \int d x~ \left| \phi^{\sigma}_i(x;t) \right|^2. \label{Eq:6}
\end{equation}
It can be shown that when $\Psi_{MB}(\vec x^A,\vec x^B;t) \to \Psi_{MF}(\vec x^A,\vec x^B;t)$ 
the corresponding natural populations obey $n_1^{\sigma}(t)=N^{\sigma}$, $n_{i\neq1}^{\sigma}(t)=0$ and then  
the first natural orbital $\phi^{\sigma}_1(x^{\sigma};t)$ reduces to the MF wavefunction 
$\phi^{\sigma}(x^{\sigma};t)$. 
Therefore, $1-n_1^{\sigma}(t)$ serves as a measure of the degree of the $\sigma$ species fragmentation \cite{mueller,penrose}.  
\begin{figure*}[t]
\includegraphics[width=0.85\textwidth]{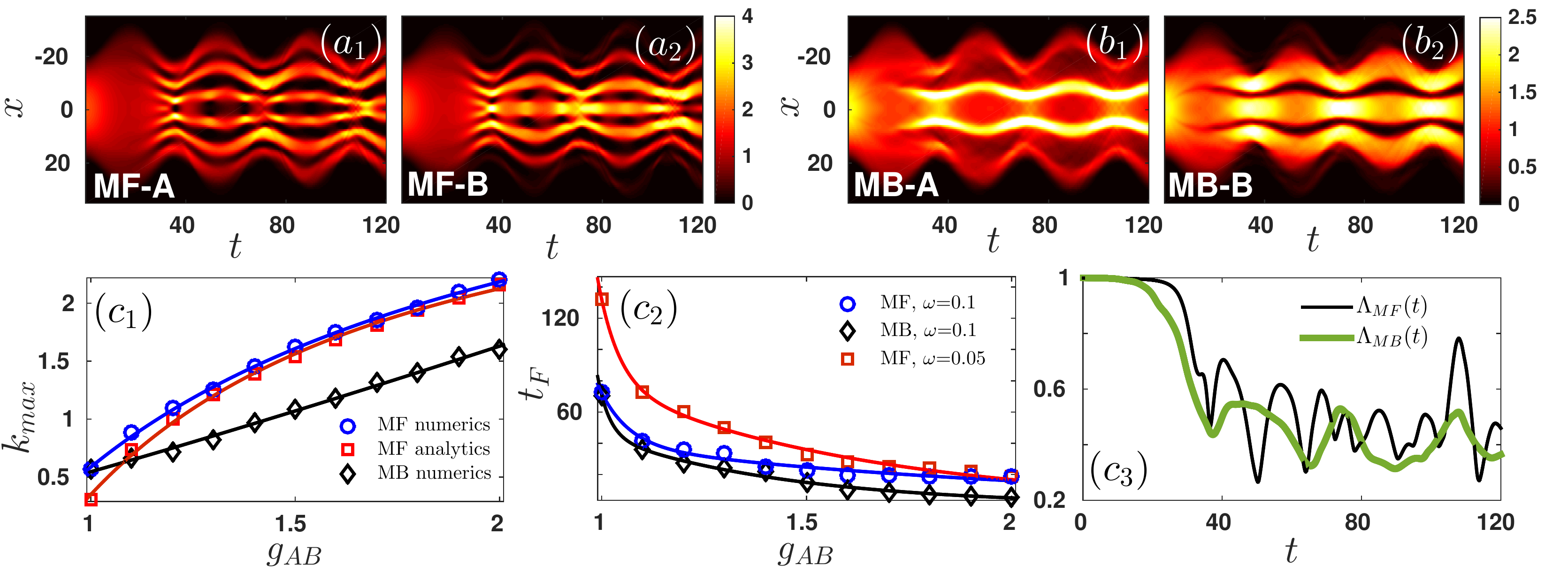}
\caption{ %(Color online) 
($a_1$), ($a_2$) [($b_1$), ($b_2$)]
  $\rho^{(1)}(x;t)$ following an interaction quench of a
binary mixture which is initially species uncorrelated, $g_{AB}=0$, 
to the immiscible phase with $g_{AB}=1.2$ for species $A$ and $B$ respectively, obtained via the
MF [MB], i.e. 1-(1,1) [15-(3,3)], approach.
($c_1$) Unstable wavenumber $k_{max}$ as a function of $g_{AB}$, and the corresponding ($c_2$) estimated time, $t_F$, for the 
filament formation (see legend). Note that solid lines in $(c_1)$ [$(c_2)$] correspond to a power-law [bi-exponential] 
fitting which is used as a guide to the eye.
($c_3$) Temporal evolution of the overlap integral calculated in the MF and the MB approach upon abruptly 
%[linearly] 
switching on the interspecies repulsion to $g_{AB}=1.2$. Both species $A$ and $B$ contain $N_A=N_B=50$ atoms  
while the trapping frequency is $\Omega=0.1$. } 
\label{Fig:1}
\end{figure*}

\section{Interaction quench dynamics}\label{quench}
In the following the quench-induced phase separation dynamics of a binary 
repulsively interacting BEC is investigated both within and beyond the MF approximation.
In particular, interspecies interaction quenches 
are performed from the miscible to the immiscible regime of interactions and vice versa. 
Recall~\cite{aochui} that species separation in the absence of a trap
occurs for $g^2_{AB}\geqslant g_{AA} g_{BB}$, while the two species  
overlap when the above inequality is not fulfilled~\cite{hall3}.  
It is relevant to note, however, that for sufficiently strong trapping--a scenario
not considered here--, the above condition is suitably modified~\cite{navarro}.
In that case, the $g_{AB}$ needed to induce immiscibility can become substantially
larger, as it needs to overcome the restoring, and hence implicitly
miscibility favoring, effect of the trap.

First we find the ground state of the system in both the MF and the MB case
for fixed intra and interspecies interactions namely $g_{AA}=1.004$, $g_{BB}=0.9544$, and  $g_{AB}=0$. 
To initialize the dynamics we then abruptly %and/or linearly 
vary the interspecies coefficient
within the interval $g_{AB}=[0, 2]$, in the dimensionless units adopted herein. 
Notice that e.g. $g_{AB}=0$ corresponds to two decoupled overlapping BECs formed around   
the center of the harmonic trap. 
With the above choice of parameters the critical point, i.e.
the miscibility-immiscibility threshold, in the absence of the trap,
is $g_{AB} \approx 0.9789$. 
The number of particles in each species is fixed to $N_A=N_B=N/2=50$,
with $N$ being the total number of particles of the system. %The dependence of the 
Dynamical phase separation for smaller bosonic ensembles is addressed in Appendix B.

\subsection{Quench dynamics to the immiscible regime}\label{quench_immicible}
\begin{figure*}[t]
\includegraphics[width=0.89\textwidth]{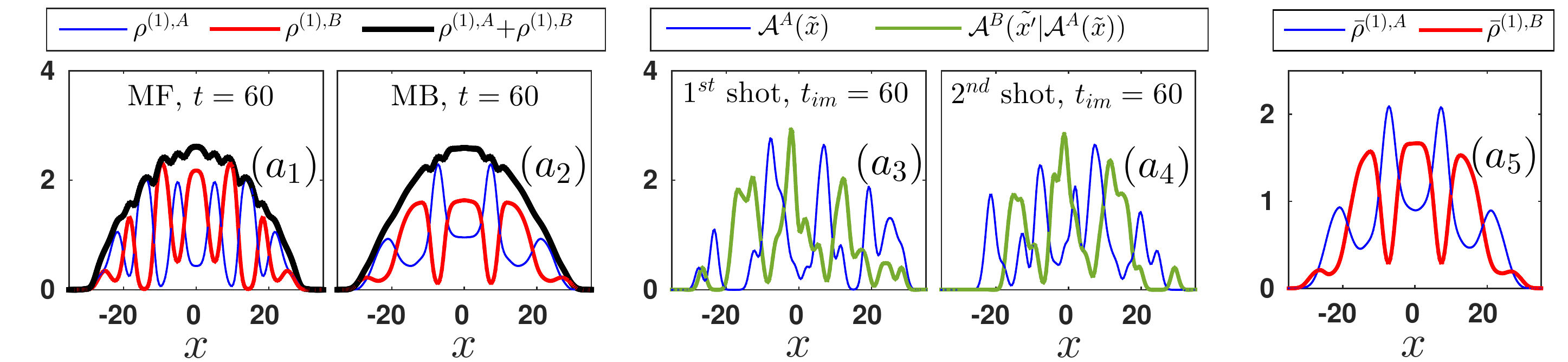}
\caption{ %(Color online) 
($a_1$), ($a_2$) Profile snapshots of the one-body density of each species $A$ and $B$, 
and the density of their sum after the filament formation within the MF and the MB case respectively
(see legend). 
($a_3$), ($a_4$) Characteristic examples of in-situ single-shot images at the MB level (see legend), 
and the corresponding averaged density ($a_5$) over 
$N_{shots}=1000$. Other parameters used are the same as in Fig.~\ref{Fig:1}.} 
\label{Fig:2}
\end{figure*}

As a first step an interaction quench of an initially species uncorrelated
(since $g_{AB}=0$)
mixture towards the immiscible regime with $g_{AB}=1.2$ is performed, driving the system abruptly 
out-of-equilibrium and letting it dynamically evolve. As shown in Fig.~\ref{Fig:1}, 
the initial ground state quickly becomes deformed
and  breaks into multiple filaments
within the MF approach, 
depicted in Figs.~\ref{Fig:1} ($a_1$) and ($a_2$),
as well as in the MB case  
shown in Figs.~\ref{Fig:1} ($b_1$) and ($b_2$). 
The dramatic phase separation observed between the two species,
and depicted for $t=60$ in the density profiles of Figs.~\ref{Fig:2} ($a_1$), ($a_2$),
results in a different number of filaments formed, the latter being greater within the MF approximation. This suggests that 
the wavenumber associated with the emergence and growth of these filaments is larger in the MF regime.
Notice that in both cases the filaments of the two species locate alternately
while the total density does not change dramatically after the filament formation.
Additionally here, the first species is found to be expelled further
off of the trap center when compared to the second species since
this configuration is energetically preferable by virtue of $g_{AA}>g_{BB}$.
Besides the filamentation of its density, each species  
performs collective oscillations that result in an expansion and contraction of the bosonic
cloud. Namely a breathing mode~\cite{Abraham,Pyzh} 
possessing a frequency $\omega_{br}=2\pi/T\approx 0.2\equiv 2\Omega$.  
Finally we remark that for a stronger post quench repulsion, $g_{AB}$, an increased number of filaments 
is observed and a more dramatic phase separation takes place, occurring much faster when compared to smaller $g_{AB}$ values.

In all cases, the dominant wavenumber associated 
with the above-observed unstable dynamics when entering the phase separated regime, 
is found to be higher in the MF approach 
when compared to the MB scenario. 
To quantify the distinct features of the manifestation of the phase separation dynamics  
within the two approaches we start by considering the stability properties 
of a homogeneous binary system of length $L$.
Within the MF approximation the spectrum of quasi-particle excitations 
consists of two branches $\Omega_{\pm}$, that in the case of equal masses between the bosons read~\cite{Tommasini} 
\begin{align}
\Omega^2_{\pm} &= \frac{k^2}{2} \left[\frac{k^2}{2} + n \left(g_{AA} + g_{BB} \vphantom{\frac{k^2}{2} + n g_{AA} + g_{BB}}\right. \right.  \nonumber \\ 
 &\qquad\qquad\qquad\!\!\!\!\!\!\!\!\!\!\left. \left. \pm  \sqrt{\left( g_{AA} - g_{BB} \right)^2 + 4 
g^2_{AB} }\right) \right], 
\label{dispersion}
\end{align}
where $n=N/2L$ denotes the linear atom density~\cite{Sabbatini}.
It turns out that if $g^2_{AB}>g_{AA} g_{BB}$, i.e. in the
immiscible regime of interactions, 
$\Omega_{-}$ becomes imaginary and  
gives rise to long wavelength modes that grow exponentially in time 
rendering the homogeneous binary system 
unstable~\cite{Timmermans}.
For $g^2_{AB}<g_{AA} g_{BB}$ both branches $\Omega^2_{\pm}$ of Eq.~(\ref{dispersion}) remain positive 
implying that the binary system is stable within this miscible regime. The two species remain then mutually overlapped 
and undergo a breathing dynamics.
Turning to $g^2_{AB}>g_{AA} g_{BB}$, the most unstable $k=k_{max}$ modes, corresponding to $\max\lbrace 
\operatorname{Im}(\Omega_{-})\rbrace$, 
are presented in Fig.~\ref{Fig:1} ($c_1$) 
for varying $g_{AB}$. 
For the numerical identification of $k_{max}$ we calculate the spectrum, 
$\tilde{\rho}^{(1)}(k;\omega)$, 
of the binary system in both the MF and the MB level. 
Among the  modes that appear in this spectrum,
we identify as the fastest growing one the mode that maximizes
the growth rate $\omega=\omega_{max}$.
As is evident in Fig.~\ref{Fig:1} ($c_1$), our numerical findings are in very good agreement with the analytical 
predictions within the MF approximation (except for very
small values of $g_{AB}$). Note here, that we have checked the validity of our
calculations for different trapping frequencies within the local density approximation (see discussion below).
However,  the unstable modes identified within the MB 
approach involve considerably shorter $k_{max}$ values which result in longer spatial scales for the filament formation (and 
thus consist of fewer filaments formed). 
For example the wavelength obtained in the MF case depicted in Fig.~\ref{Fig:1} ($c_1$) for $g_{AB}=1.2$
is $\lambda_{MF}=2\pi/k_{max} \approx 5.76$ ($k_{max}\approx 1.09$) while at the MB level we get the value 
$\lambda_{MB}\approx 8.73$ ($k_{max}\approx 0.72$).
The observed difference of $k_{max}$ between the MF and MB evolution can be attributed to the 
participation of additional MB excitations which lie beyond the linear response theory as demonstrated, e.g., in 
\cite{Grond_lr} for single component setups.  

Additionally, having identified the wavenumber associated with
the fastest growth,
we can also infer the time at which the filament formation
occurs. We have estimated this time, namely $t=t_{F}$, by 
identifying the time at which the amplitude of this wavenumber, $k=k_{max}$, starts to grow. 
The formation time, $t_F$, 
is illustrated in Fig.~\ref{Fig:1} ($c_2$) for increasing $g_{AB}$ and is fitted by a   
biexponential function. 
It is evident that close to the miscibility-immiscibility threshold 
($g_{AB}\approx 1$) both approaches coincide, while deviations between the two become apparent as we increase the 
interspecies interactions. Note also that decreasing the trapping strength towards the homogeneous case alters 
the time scale at which the instability manifests itself, the more, the closest we are to the above threshold. 

To quantify the degree of phase separation we evaluate the overlap 
integral~\cite{jain,Bandyopadhyay}
\begin{eqnarray}
\Lambda(t)=\frac{\left[\int dx \rho ^{(1),A}(x;t)  \rho ^{(1),B}(x;t)  \right]^2}{\left[\int dx \left(\rho ^{(1),A}(x;t) 
\right)^2  \right] 
\left[\int dx \left(\rho ^{(1),B} (x;t) \right)^2  \right]},
\label{overlap}
\end{eqnarray}
where, $\Lambda(t)=1$ [$\Lambda(t)=0$] denotes complete [zero] overlap of the two species 
upon abruptly driving the system out-of-equilibrium. 
As depicted in Fig.~\ref{Fig:1} $(c_3)$ the transition to immiscibility is signalled at slightly earlier times
in the MB approach with the overlap between the two species being of about $50\%$ on average, while being 
almost $60\%$ on average within the MF approximation. Moreover, the abrupt quench protocol entails rapid oscillations 
in the MF case when compared to the smoother drop down towards immiscibility observed in the MB 
scenario. It is worth mentioning at this point, that the same overall phenomenology is observed even upon linearly 
quenching the system between the same initial and final $g_{AB}$ values (results not shown here for brevity). 
The key outcome in this case is that the 
filamentation process is signalled at times proportional to the ramping time used resulting to a larger $\Lambda (t)$ when
compared to the abrupt quench protocol.
\begin{figure*}[ht]
\includegraphics[width=1.0\textwidth]{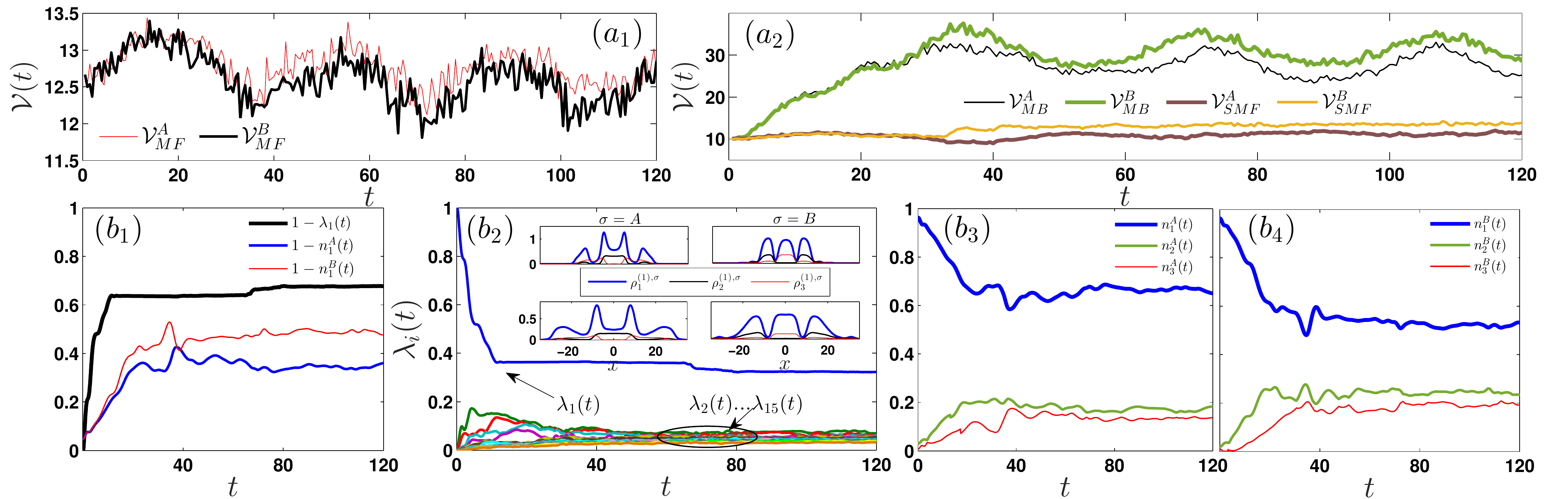}
\caption{ %(Color online) 
Temporal evolution of the variance, $\mathcal{V}(t)$, obtained via in-situ single-shot
measurements in ($a_1$) the MF, ($a_2$) the SMF (see text for the relevant explanation)
and the 	MB case respectively (see legends).
($b_1$) Deviation from unity of the first natural species population and the first natural population
of the $\sigma$-species respectively (see legend).
($b_2$) Evolution of the natural species populations, $\lambda_i(t)$ with $i=1, \ldots,15$. 
Insets illustrate snapshots, at $t=36$ (top panels), and $t=55$ (bottom panels),
during propagation of the first three modes of entanglement, $\rho^{(1),\sigma}_k$ ($k=1,2,3$) for the $\sigma=A, B$ species.
($b_3$), ($b_4$) Temporal evolution of the natural populations $n_i(t)$ for species $A$ and $B$ respectively.
In all cases the system is quenched from $g_{AB}=0$ to $g_{AB}=1.2$, while other parameters used
are the same as in Fig. \ref{Fig:1}.} 
\label{Fig:3}
\end{figure*}

\subsection{Single-shot simulations}

As a next step we elaborate on how the MB character of the dynamics  
can be inferred by performing in-situ single-shot absorption measurements \cite{kaspar}. 
Such measurements probe the spatial configuration of the atoms which is dictated by the MB probability distribution.      
An experimental image refers to a convolution of the spatial particle configuration with a point spread function. 
The latter describes the response of the imaging system to a point-like absorber (atom). 
Relying on the MB wavefunction being available within ML-MCTDHB we mimic the above-mentioned experimental 
procedure and simulate such single shot images for both species A [namely $\mathcal{A}^A(\tilde{x};t_{im})$] 
and species B [i.e. $\mathcal{A}^B(\tilde{x'}|\mathcal{A}^A(\tilde{x});t_{im})$] at each instant of the evolution (for more 
details see Appendix A) when we consecutively image first the $A$ and then the $B$ species.  
We remark that the employed point spread function (being related to the experimental resolution), 
consists of a Gaussian possessing a width $w=1 \ll l\approx 3.2$. %where $l$ denotes the harmonic oscillator length. 

Figs. \ref{Fig:2} ($a_3$), ($a_4$) illustrate the first and the second simulated in-situ single-shot images 
at $t_{im}=60$ for both species, namely 
$\mathcal{A}^A(\tilde{x};t_{im}=60)$, and $\mathcal{A}^B(\tilde{x'}|\mathcal{A}^A(\tilde{x});t_{im}=60)$.  
It is evident that in both shots the two species exhibit a phase separated behavior resembling this way the overall tendency 
observed in the one-body density [see also Fig. \ref{Fig:2} ($a_2$)]. 
However, a direct observation of the one-body density in a single-shot image is 
not possible due to the small particle number of the considered 
binary bosonic gas, $N_A=N_B=50$, as well as the presence of multiple orbitals in the system.
The MB state builds upon a superposition of multiple orbitals [see Eqs.~(\ref{Eq:4})-(\ref{Eq:5})]
and therefore imaging an atom alters the MB state of the remaining atoms and hence their one-body
density. This is in direct contrast to a MF product state, composed from a single macroscopic orbital, 
where the imaging of an atom does not affect the distribution of the rest  
(see also the discussion below for the corresponding variance). 
Note also here that the above-mentioned single-shot images are reminiscent of the experimental images obtained in 
a two-dimensional (2D) geometry when examining the phase separation process~\cite{wieman}. 
To reproduce the one-body density of the system one needs to rely on an average 
of several single-shot images.  
Indeed, Fig. \ref{Fig:2} ($a_5$) shows within the MB approach the obtained average, $\bar{\rho}^{(1),\sigma}$, 
over $N_{shots}=1000$ images
for both species, namely $\bar{\mathcal{A}}^A(\tilde{x};t_{im})=1/N_{shots}\sum_{k=1}^{N_{shots}} 
\mathcal{A}_k^A(\tilde{x};t_{im})$ 
and $\bar{\mathcal{A}}^B(\tilde{x}^{'}|\mathcal{A}^A(\tilde{x});t_{im})=1/N_{shots}\sum_{k=1}^{N_{shots}} 
\mathcal{A}_k^B(\tilde{x}^{'}|\mathcal{A}^A(\tilde{x});t_{im})$ respectively. 
As expected, a direct comparison of this averaging and the actual one-body density obtained within the MB approach 
[see Figs. \ref{Fig:2} ($a_2$) and ($a_5$)] reveals that they are almost identical. 
Finally, let us remark here that similar observations can be made when performing the single-shot procedure 
initially for the $B$ and then for the $A$ species.  

Let us now investigate whether the presence of correlations
can be deduced from the time evolution of the variance $\mathcal{V}(t)$ of a sample of single-shot 
measurements~\cite{Lode,c6,zorzetos}. 
As before, we mainly focus on the scenario where the imaging is performed first on the $A$ and then on the $B$ species, but 
the same results can be obtained for the reverse consecutive imaging process. 
The variance of a set of single-shot measurements $\{\mathcal{A}_k^A(\tilde{x})\}_{k=1}^{N_{shots}}$ concerning the $A$ 
species reads 
\begin{equation}
\begin{split}
&\mathcal{V}^A(t_{im})=\\&\int d\tilde{x} \frac{1}{N_{shots}} \sum_{k=1}^{N_{shots}} [\mathcal{A}_k^A(\tilde{x};t_{im})-
\bar{\mathcal{A}}_k^A(\tilde{x};t_{im})]^2.  
\end{split}\label{Eq:7}
\end{equation}
In the same manner, one defines the variance of a set of single-shots $\{\mathcal{A}_k^B(\tilde{x}^{'}|
\mathcal{A}^A(\tilde{x}))\}_{k=1}^{N_{shots}}$ referring to the $B$ species  
\begin{equation}
\begin{split}
&\mathcal{V}^B(t_{im})=\int d\tilde{x}' \frac{1}{N_{shots}}\times \\& \sum_{k=1}^{N_{shots}} [\mathcal{A}_k^B(\tilde{x}'|
\mathcal{A}_k^A(\tilde{x});t_{im})-
\bar{\mathcal{A}}_k^B(\tilde{x}'|\mathcal{A}_k^A(\tilde{x});t_{im})]^2.   
\end{split} \label{Eq:8}
\end{equation}
Figs.~\ref{Fig:3} ($a_1$), ($a_2$) present both $\mathcal{V}^A(t)$ and $\mathcal{V}^B(t)$ with $w=1$, and $N_{shots}=1000$ at 
the MF and the MB level respectively.  
As it can be seen, at the MF approximation $\mathcal{V}_{MF}^A(t)$ and $\mathcal{V}_{MF}^B(t)$ remain almost constant 
exhibiting small amplitude oscillations which essentially resemble the breathing motion that both species feature. 
However, when inter and intraspecies correlations are taken into account $\mathcal{V}_{MB}^A(t)$ and $\mathcal{V}_{MB}^B(t)$ 
show a completely different behavior. 
In particular, an increasing tendency is observed at the initial stages of the unstable dynamics,
while after the filament formation ($t_F\approx 27$), $\mathcal{V}_{MB}^{\sigma}(t)$
undergoes large amplitude oscillations reflecting the global breathing of each bosonic cloud.  
More importantly, the aforementioned increasing tendency of the variance resembles the growth rate of the 
entanglement, [see $1-\lambda_1(t)$ in Fig. \ref{Fig:3} ($b_1$)] and the corresponding discussion below]. 
The above resemblance can be explained as follows.  
In a perfect condensate, i.e. $\lambda_1(t)=1$ and $n_1^{\sigma}(t)=1$, 
$\mathcal{V}_{MF}^{\sigma}(t)$ is almost constant during the dynamics  
as all the atoms in the corresponding single-shot measurement are picked from the same SPF $\varphi^{\sigma}(t)$ 
[see also Eq. (\ref{Eq:2})]. 
The only relevant information that is imprinted in $\mathcal{V}_{MF}^{\sigma}(t)$ 
concerns the global motion, here the breathing mode, of the entire cloud. 
It is also worth mentioning here that $\mathcal{V}_{MF}^A(t)\approx \mathcal{V}_{MF}^B(t)$ during the MF evolution, testifying 
the absence of both inter and intraspecies correlations.  
The observed negligible differences between $\mathcal{V}_{MF}^A(t)$, and $\mathcal{V}_{MF}^B(t)$ 
[hardly visible in Fig. \ref{Fig:3} ($a_1$)] are 
caused by the slight deviations in the magnitude of the breathing motion that each species undergoes. 

On the contrary, for a MB system where entanglement and fragmentation are present due to the inclusion of inter and 
intraspecies correlations, 
the corresponding MB state consists of an admixture of various mutually orthonormal species functions 
$\Psi_k^A(t)$ and $\Psi_k^B(t)$ respectively,  $k=1,2,...,15$ [see Eq. (\ref{Eq:3})] each of them building upon different 
mutually orthonormal SPFs $\varphi_i^A(t)$ and $\varphi_i^B(t)$ respectively,  $i=1,2,3$ [see also Eq. (\ref{Eq:4})].  
In this way, the corresponding single-shot variance is drastically altered from its MF counterpart as the atoms are picked 
from the above-mentioned superposition and thus their distribution in the cloud depends strongly on the 
position of the already imaged atoms \cite{kaspar,Lode,filled_vortex}, see also Appendix A.    
To fairly discern between the impact of the inter and intraspecies correlations on the variance we first inspect 
$\mathcal{V}^{\sigma}(t)$ when neglecting the entanglement between the species [this approach will be referred in the 
following as species mean-field approximation (SMF)]. 
Namely we calculate $\mathcal{V}_{SMF}^{\sigma}(t)$ assuming that the $N_{\sigma}$-body state of each species is described by 
only one species function ($\Psi_k^A(t)=\Psi_k^B(t)$=0 for $k\neq1$) that builds 
upon distinct SPFs $\varphi_i^A(t)$ and $\varphi_i^B(t)$,  $i=1,2,3$. 
As shown in Fig. \ref{Fig:3} ($a_2$) during the filamentation process $\mathcal{V}_{SMF}^{\sigma}(t)$ increases slightly and 
$\mathcal{V}_{SMF}^{A}(t)\approx\mathcal{V}_{SMF}^{B}(t)$ while 
at later time instants $\mathcal{V}_{SMF}^{A}(t)<\mathcal{V}_{SMF}^{B}(t)$.  
This latter deviation is attributed to the different degree of fragmentation 
[$1-n_1^{\sigma}(t)$, see e.g. Fig. \ref{Fig:3} ($b_1$)] 
that each species possesses after the filamentation process $t>27$.  
Having identified that the presence of fragmentation essentially causes a slight increase on the single-shot variance 
and more importantly gives rise to deviations between the $\mathcal{V}_{SMF}^{\sigma}(t)$'s of the two species we can 
elaborate on 
the impact of the entanglement when also interspecies correlations are taken into account. 
In the MB case $\mathcal{V}_{MB}^{\sigma}(t)$ shows a remarkable increasing tendency during the filamentation process 
highlighting this way the presence of entanglement in the system. 
Indeed, the increase of entanglement [evident in $1-\lambda_1^{\sigma}(t)$] and consequently of the variance can be attributed 
to the build up of higher-order superpositions during the filamentation process. 
Since the absorption imaging destroys the entanglement between the species, we expect that the single-shot images heavily 
depend on the first few imaged atoms giving rise to pronounced $\mathcal{V}_{MB}^{\sigma}(t)$.  
We further remark that this increasing tendency of the variance becomes more pronounced (reduced) for larger (smaller) 
quench values (results not included for brevity). 
Moreover, during the filamentation process $\mathcal{V}_{MB}^A(t)\approx\mathcal{V}_{MB}^B(t)$ but after their formation 
$\mathcal{V}_{MB}^A(t)<\mathcal{V}_{MB}^B(t)$. 
This latter deviation can be attributed to the different degree of fragmentation that 
builds up during evolution in each of the two species 
[compare $1-n_1^{\sigma}(t)$ for $t\ge40$ illustrated in Fig. \ref{Fig:3} 
($b_1$)]. 
We finally note that the above-described overall increasing behavior of 
$\mathcal{V}_{MB}^A(t)$ and $\mathcal{V}_{MB}^B(t)$ is robust also 
for smaller samplings of single-shot measurements, e.g. $N_{shots}=100$, or different widths, e.g. $w=0.5$,
(results not shown here for brevity). 
\begin{figure*}[ht]
\includegraphics[width=0.9\textwidth]{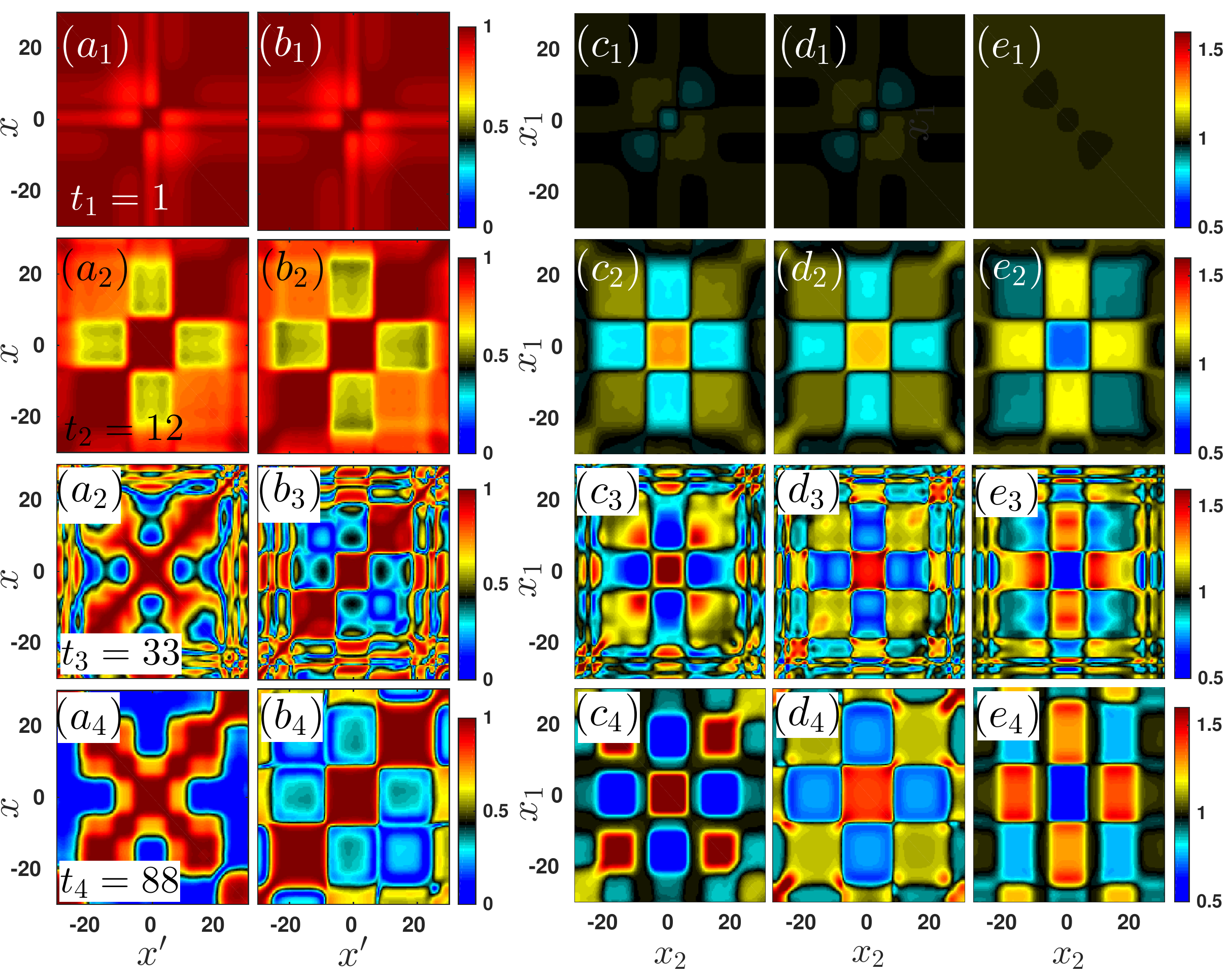}
\caption{ %(Color online) 
($a_1$)-($a_4$) [($b_1$)-($b_4$)] One-body normalized correlation function $|g^{(1),A}(x,x';t)|$ [$|g^{(1),B}(x,x';t)|$] 
shown for different time instants during the evolution.   
($c_1$)-($c_4$) [($d_1$)-($d_4$)] Snapshots of the two-body correlation function $|g^{(2),A}(x_1,x_2';t)|$ [$|g^{(2),B}
(x_1,x_2';t)|$] of the $A$ [$B$] species.  
($e_1$)-($e_4$) Interspecies two-body correlation function $|g^{(2),(A,B)}(x_1,x_2';t)|$.
In all cases the same selected time instants during propagation are illustrated (see legends).
The remaining parameter values are the same as in Fig. \ref{Fig:1}.} 
\label{Fig:4}
\end{figure*}
\subsection{Correlation dynamics}
The degree of entanglement is encoded in the species functions of the binary system, 
i.e. $\Psi_k^{\sigma}(\vec x^{\sigma};t)$, with $\sigma=A,B$, being weighted by the $\lambda_k(t)$ coefficients. 
We remind the reader that if $\lambda_1(t)=1$ and $\lambda_i(t)=0$ ($i=2,...,k$) then the non-entangled limit is reached 
while if $\lambda_k(t)\neq 0$ the more modes are occupied the more
strongly entangled the binary system is~\cite{note_ent}.
In particular, by considering the evolution of the natural 
occupations $\lambda_k(t)$, depicted in Fig.~\ref{Fig:3} $(b_2)$ it is observed that from the beginning of the quench 
induced dynamics the occupation of the initial single mode (non-entangled) wavefunction 
reduces rapidly and higher-lying modes become 
spontaneously populated.  
Notice that before the filament formation, e.g. at $t\approx 13$, $\lambda_1\approx 0.37$    
and $\lambda_2\approx\lambda_3\approx 0.12$, while after the breaking ($t\approx 27$) 
the amplitude of the higher-lying modes drops below $0.1$ and remains
in this ballpark till the end of the propagation.
The insets depict selected time instants during the phase separation process of the first three modes of entanglement:  
namely, just after the breaking [upper insets in Fig.~\ref{Fig:3} $(b_2)$] 
and the consequent filamentation of the MB wavefunction, and for larger propagation times 
[lower insets in Fig.~\ref{Fig:3} $(b_2)$] 
corresponding to $\Lambda(t)\approx 0.5$ 
during evolution [see also Fig.~\ref{Fig:1} ($c_3$)].
In all cases the leading order mode weighted by $\lambda_1$, and the first two 
of the higher-lying modes that are predominantly occupied, 
weighted by $\lambda_2$ and $\lambda_3$ respectively, are shown for both the $A$ and $B$ species. 
As it is evident, the dominant mode clearly captures all the filaments formed for both species. 
The second mode for species $A$ builds a hump at the location centered around the density dip of the
first mode, while it also follows the outer filaments formed, and the corresponding third mode  
mostly supports the inner filaments. 
As far as the B species is concerned the above observed phenomenology is somewhat reversed. Notice that, 
the second mode mostly follows the outer filaments, and the third mode is found to be predominantly 
associated with the filaments developed closer to the trap center. 

To further elaborate on the MB nature of the observed quench dynamics
we next examine the population of the natural orbitals shown in Figs.~\ref{Fig:3} ($b_3$), ($b_4$). 
The occupations of the three natural orbitals used for each of the two species are significant
from the early stages of the dynamics, with the two lower-lying orbitals being monotonically ordered, 
acquiring lower populations during evolution. 

As already discussed in Sec. II the non-negligible population of both $\lambda_k$
and $n^{\sigma}_k$ ($k>1$) signifies the presence of inter- and intraspecies correlations respectively. 
To identify the degree of intraspecies correlations at the one-body level during the quench dynamics, we employ the 
normalized spatial first order correlation function \cite{Naraschewski,density_matrix}  
\begin{equation}
g^{(1),\sigma}(x,x';t)=\frac{\rho^{(1),\sigma}(x,x';t)}{\sqrt{\rho^{(1),\sigma}(x;t)\rho^{(1),\sigma}(x';t)}}. \label{Eq:9}
\end{equation} 
This quantity measures essentially the proximity of the MB state to a MF (product) state for a fixed set of coordinates $x$, 
$x'$.  
$\rho^{(1),\sigma}(x,x';t)$ is the one-body reduced density matrix of the $\sigma$ species [see also Eq. (\ref{Eq:5})] and  
$\rho^{(1),\sigma}(x;t)\equiv \rho^{(1),\sigma}(x,x'=x;t)$.
Furthermore, $|g^{(1),\sigma}(x,x';t)|$ takes values within the range $[0,1]$.  
Note that, two different spatial regions $R$, $R'$, with $R \cap R' = \varnothing$, exhibiting   
$|g^{(1),\sigma}(x,x';t)|= 0$, $x\in R$, $x'\in R'$ ($|g^{(1),\sigma}(x,x';t)|= 1$, $x\in R$, $x'\in R'$) are referred to 
as fully incoherent (coherent). 
The absence of one-body correlations in the condensate is indicated by $|g^{(1),\sigma}(x,x';t)|=1$ for every $x$, $x'$ while 
the case that at least two distinct spatial regions are partially incoherent i.e. $|g^{(1),\sigma}(x,x';t)|<1$ signifies the 
emergence of correlations. 

Figs. \ref{Fig:4} ($a_1$)-($a_4$) [($b_1$)-($b_4$)] present $|g^{(1),A}(x,x';t)|$ [$|g^{(1),B}(x,x';t)|$] for different 
time instants during the dynamics, namely before and after the filamentation process. 
At initial time instants [see Figs. \ref{Fig:4} ($a_1$),($a_2$) and ($b_1$), ($b_2$)] where the density deformation sets in, 
one-body correlations begin to develop.  
For instance $|g^{(1),\sigma}(x,x';t)|\approx 0.5$ between the central and the outer BEC regions 
(in which the filaments are formed later on, see e.g. at $x\approx0$, $x'\approx15$ at $t=12$), while $|g^{(1),\sigma}
(x,x';t)|\approx 0.8$  
among the outer regions ($x=-x'\approx15$ at $t=12$).    
An augmented character of $|g^{(1),\sigma}(x,x';t)|$ for increasing distances (e.g. for fixed $x\approx0$, towards 
$x'\approx25$ at $t=7$) 
is also observed.
For later evolution times, i.e. after the filamentation process, a significant build up of one-body correlations occurs for 
both species. 
Referring to $|g^{(1),A}(x,x';t)|$, see Figs. \ref{Fig:4} ($a_3$), ($a_4$), we observe that each filament is perfectly 
coherent with itself (see the diagonal elements), while a small amount of correlations occurs between the inner filaments ($|
g^{(1),A}(x\approx6,x'\approx-6;t=33)|\approx0.9$) 
or the outer ones ($|g^{(1),A}(x\approx14,x'\approx-14;t=33)|\approx0.8$). 
More importantly, strong correlations appear between neighbouring inner and outer filaments 
as well as among an inner (outer) 
filament and its long distance outer (inner) one ($|g^{(1),A}(x,x';t)|\approx0.5$) signalling their independent nature. 
Finally, significant losses of coherence are observed between the inner (outer) filaments and the central dip.   
Turning to $|g^{(1),B}(x,x';t)|$, see Figs. \ref{Fig:4} ($b_3$), ($b_4$), it is evident that strong correlations appear 
among each outer and the central filaments (see e.g. $x\approx10$, $x'\approx0$ at $t=33$) as well as between the outer ones 
($x=-x'\approx10$ at $t=33$).   
This latter behavior is manifested by the almost vanishing 
off-diagonal elements of $|g^{(1),B}(x,x';t)|$ after the 
filamentation process, indicating a tendency of localization of each filament formed. 
\begin{figure*}[ht]
\includegraphics[width=0.98\textwidth]{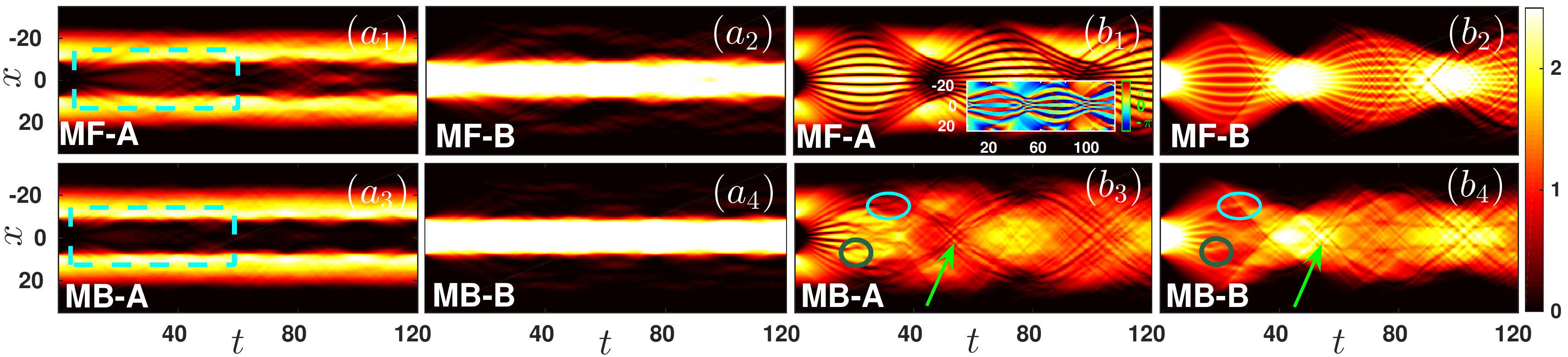}
\caption{ %(Color online) 
($a_1$), ($a_2$) [($a_3$), ($a_4$)] 
Quenched $\rho^{(1),A}(x;t)$, and $\rho^{(1),B}(x;t)$ for an immiscible ($g_{AB}=1.4$) to an immiscible 
($g_{AB}=1.0$) transition, within the MF [MB] approach.  
($b_1$), ($b_2$) [($b_3$), ($b_4$)] The same as the above but for an immiscible ($g_{AB}=1.4$) to a miscible 
($g_{AB}=0.5$) transition. The inset in $(b_1)$ shows the corresponding evolution of the phase $\arg[\phi^{(A}(x^A_i;t)]$ in the MF case.
Other parameters used are the same as in Fig.~\ref{Fig:1}.}
\label{Fig:5}
\end{figure*}

Having discussed in detail the significance of one-body intraspecies correlations, we next
quantify the degree of second order intra- and interspecies correlations by inspecting the normalized two-body correlation 
function~\cite{density_matrix}  
\begin{equation}
\begin{split}
g^{(2),\sigma \sigma'}(x_1,x_2;t)=\frac{\rho^{(2),\sigma \sigma'}(x_1,x_2;t)}{\rho^{(1),\sigma}(x_1;t)\rho^{(1),\sigma'}
(x_2;t)}.\label{Eq:10}
\end{split}
\end{equation} 
$\rho^{(2),\sigma \sigma'}(x_1,x_2;t)=\bra{\Psi_{MB}(t)}\Psi^{\dagger,\sigma}(x_1)\Psi^{\dagger,\sigma'}
(x_2) \\
\Psi^{\sigma}(x_1)\Psi^{\sigma'}(x_2)\ket{\Psi_{MB}(t)}$ 
is the diagonal two-body reduced density matrix referring to the probability of measuring two particles located at positions 
$x_1$, $x_2$ at time $t$.
$\Psi^{\dagger,\sigma}(x_i)$ [$\Psi^{\sigma}(x_i)$] is the bosonic field operator that creates (annihilates) a $\sigma$ 
species boson at position $x_i$. 
Regarding the same (different) species, i.e. $\sigma=\sigma'$ ($\sigma \not=\sigma'$),  
$|g^{(2),\sigma \sigma'}(x_1,x_2;t)|$ accounts for 
the intraspecies (interspecies) two-body correlations 
and is also experimentally accessible via in-situ density density fluctuation measurements \cite{granulation,Tavares,Endres}.   
We remark here that a perfectly condensed MB state leads to $|g^{(2),\sigma \sigma'}(x_1,x_2;t)|=1$ and it is termed fully 
second order coherent or uncorrelated. 
However, if $|g^{(2),\sigma \sigma'}(x_1,x_2;t)|$ takes values smaller (larger) than unity the state is referred to as 
anti-correlated (correlated).  

Let us first comment on the intraspecies two-body correlated character of the dynamics.
Focusing on $|g^{(2),AA}(x_1,x_2;t)|$ we observe a consecutive formation of two-body correlations during the dynamics, see 
Figs. \ref{Fig:4} ($c_1$)-($c_4$).  
Besides a bunching tendency (smaller for the inner filaments) of two bosons to lie within each filament (see the diagonal 
elements), a correlated behavior 
is observed among two parity symmetric outer ones (see e.g. $x_1=-x_2=14$ at $t=33$).  
In addition, an outer filament is anti-correlated both with an inner one ($x_1\approx14$, $x_2\approx6$ at $t=33$) as well as 
with the central dip ($x_1\approx14$, $x_2\approx0$). 
Combining this latter behavior with the above suppression of $|g^{(1),A}(x,x^{'};t)|$
between the filaments, implies the formation of domain-wall-like structures between the area of central filaments and 
an outer one. 
Another interesting observation here is that the region between neighbouring inner and outer filaments (e.g. $x_1\approx16$ at 
$t=33$) is strongly correlated 
(anti-correlated) with its parity symmetric one.    
Similar observations can also be made for the $|g^{(2),BB}(x_1,x_2;t)|$, see Figs. \ref{Fig:4} ($d_1$)-($d_4$).  
Evidently, it is preferable for two bosons to reside within each filament (see the diagonals) or one in each of the outer 
filaments (e.g. $x_1=-x_2\approx10$, $t=33$).  
The central filament is anti-correlated with the outers %($x_1\approx0$, $x_2\approx9$ at $t=33$) 
throughout the dynamics and since $|g^{(1),BB}(x,x';t)|\rightarrow 0$ in the same region, 
the formation of a domain-wall-like structure between a 
central and an outer filament can be 
inferred.  

As a next step we inspect the interspecies correlation dynamics via $|g^{(2),AB}(x_1,x_2;t)|$, 
see Figs. \ref{Fig:4} ($e_1$)-($e_4$).  
Here, an outer A species filament ($x_1\approx14$ at $t=33$) is anti-correlated (correlated) with the corresponding B species 
outer located at $x_2\approx10$ (central at $x_2=0$).   
However, an inner A species filament ($x_1\approx5$ at $t=33$) is correlated (anti-correlated) with the respective B species 
outer (central) one. 
Moreover, we find that the central dip of the A species exhibits a correlated (anti-correlated) behavior with the outer 
(central) B species filaments. 
Summarizing the outcome of $|g^{(2),AB}(x_1,x_2;t)|$ is two-fold. The fact $|g^{(2),AB}(x_1,x_2;t)|\neq 1$
indicates the entangled character of the MB binary system. Additionally, the presence of anti-correlations
between the inner and outer filaments of $A$ and $B$ species respectively (or vice versa)
supports the phase separation process being imprinted as domain-walls at the two-body level.       
\begin{figure*}[tbp]
\includegraphics[width=0.98\textwidth]{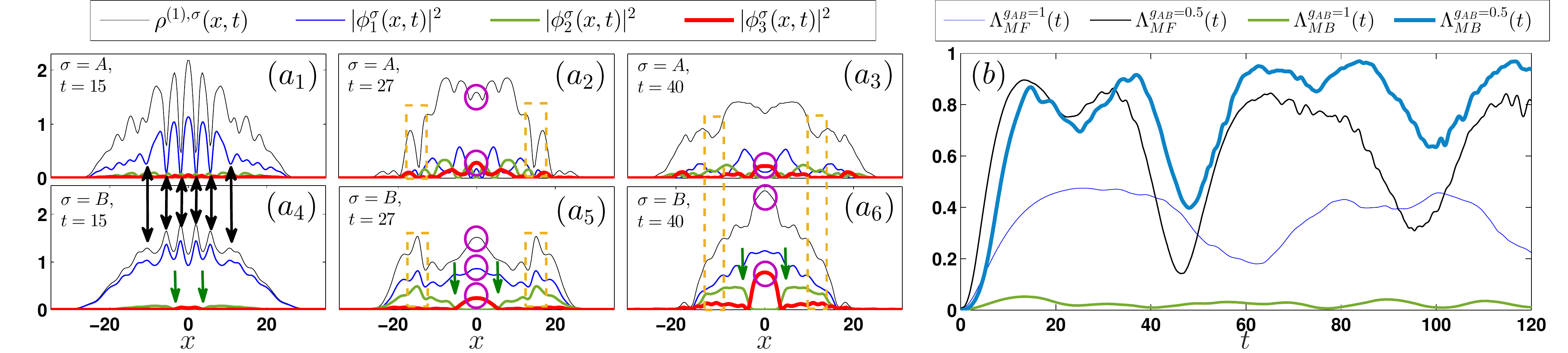}
\caption{ %(Color online)  
($a_1$)-($a_6$) Profile snapshots of the one-body density of each species $A$ and $B$, 
as well as of the three natural orbitals used for each species for the negative quench scenario (see legend). 
We also note that the second and the third natural orbitals 
of species $B$ are multiplied by a factor of 8 to provide better visibility.  
($b$) Temporal evolution of the overlap integral calculated in both approaches and for both  
transitions depicted in Fig.~\ref{Fig:5} (see legend). 
Other parameters used are the same as in Fig.~\ref{Fig:1}.}
\label{Fig:6}
\end{figure*}

\subsection{Reverse quench dynamics}\label{reverse_quench}
Up to now we explored cases which involve transitions from the miscible to the immiscible phase, 
by initializing the dynamics from the species 
uncorrelated ($g_{AB}=0$) case and abruptly switching on the interspecies repulsion. 
Our aim here, is to consider the reverse process, namely initialize the system from a species correlated ground state 
with $g_{AB}=1.4$, i.e. deep in the immiscible regime of interactions, and suddenly reduce $g_{AB}$. 
A characteristic example of an immiscible to the immiscible transition with post-quench value $g_{AB}=1.0$
is realized in Figs.~\ref{Fig:5} ($a_1$)-($a_4$).    
Notice that the phase separated species remain as such at all times with species $A$ forming two humps 
symmetrically placed around the center of the trap. Closer inspection of
the central almost zero density region,
suggests that two hardly visible density dips are spontaneously formed
in the regions indicated by dashed rectangles in Figs.~\ref{Fig:5} ($a_1$)
and ($a_3$) for the MF and the MB case respectively. 
These density dips interact with the density peaks created in this species right at their phase boundary, and
via this interaction multiple interference fringes can be seen around the center of the trap in both approaches.
It is these events which are more pronounced in 
the MF than in the MB approach, that result in the differences measured in the overlap between the two species.  
In particular as shown in Fig.~\ref{Fig:6} $(b)$, $\Lambda_{MF}(t)\approx 0.35$ on average, while
$\Lambda_{MB}(t)\lesssim 0.05$ during evolution, which is significantly smaller. 
The location of these dips is 
also the location of a ``giant'' density hump formed in species $B$. 
It is also worth mentioning at this point that the evolved phase separated state formed here, 
consists the 1D analogue of the so-called ``ball'' and ``shell''
state that forms in higher-dimensional binary BECs~\cite{usadsh}.  
\begin{figure*}[tbp]
\includegraphics[width=1.0\textwidth]{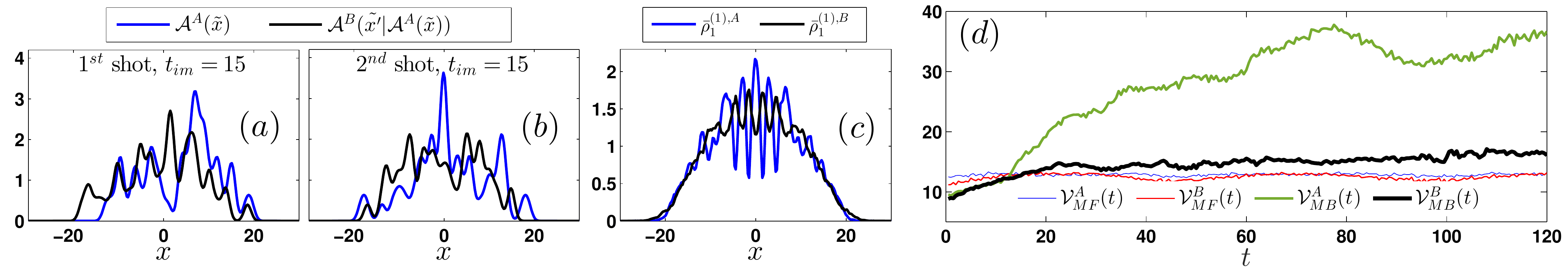}
\caption{ %(Color online)  
($a$), ($b$) Characteristic examples of in-situ single-shot images at the MB level (see legend), 
and the corresponding averaged density ($c$) over 
$N_{shots}=1000$. ($d$) Temporal evolution of the variance, $\mathcal{V}(t)$, obtained via in-situ single-shot measurements 
in both approaches (see legend).
Other parameters used are the same as in Fig.~\ref{Fig:1}.}
\label{Fig:7}
\end{figure*}

However a far more rich dynamical behavior of the binary system is observed when the two immiscible 
species are abruptly quenched towards the miscible 
regime, with the post-quench value $g_{AB}=0.5$.  
Such a situation is illustrated in Figs.~\ref{Fig:5} ($b_1$),($b_2$) [($b_3$), ($b_4$)] within the MF [MB] approach.  
The quench dynamics leads to the formation of multiple DAD solitary waves~\cite{Danaila,KevrekidisDAD}  
both in the MF and in the MB approach. 
In the former case, the DAD structures are directly discernible 
and can be seen to interact and perform oscillations, splitting
and recombining within the parabolic trap, in a way reminiscent
of the one-component
dark solitons in the experiments of~\cite{markus1,markus2}.
To verify the nature of these structures we further depict as an inset in 
Fig.~\ref{Fig:5} ($b_1$) the spatio-temporal evolution of the phase, where the phase jumps 
corresponding to the location of each dark soliton shown in the density can be easily seen.  
In contrast to that, in the MB scenario the dynamical evolution of these DAD  
structures is less transparent, since the system in this case is strongly correlated and the background 
at which the solitons are formed is highly excited. 
Recall that dark-bright states are prone to decay in the presence of quantum fluctuations~\cite{lgs} into faster 
(travelling towards the periphery of the cloud) and slower (remaining closer to the trap center) 
solitary waves. 
A similar dynamical phenomenology is also observed here for the above-mentioned DAD states. 
Indeed, at the early stages of the dynamics several decay events 
occur. Two case examples of such a decay are marked with circles in Figs.~\ref{Fig:5} ($b_3$), ($b_4$)
corresponding to an initially fast and an initially 
slow DAD pair respectively. This way in the MB case the number of the solitary waves formed increases 
when compared to the initial stages of the dynamics and thus multiple collision events occur during propagation.  
We can clearly distinguish a collision event  
closer to the trap center at $t_F^{MB}\approx 27$ which results to a merger. 
On the other hand, the corresponding fast moving DAD states reach at 
different times the periphery of the cloud and thus multiple 
collision events occur at different times during evolution. 
A case example of such a collision is indicated with arrows in   
Figs.~\ref{Fig:5} ($b_3$), ($b_4$). 

To expose the multi-orbital nature of the above dynamics, both the one-body density as 
well as the different orbital contributions are depicted in Figs.~\ref{Fig:6} ($a_1$)-($a_6$) at initial ($t=15$), 
intermediate ($t=27$) and larger evolution times ($t=40$). 
Notice that at initial times the two species are still phase separated, while 
the first orbital predominantly describes the MB 
dynamics of the system. Here, we can easily measure the number of DAD solitary waves that are initially formed, 
illustrated with two-directional arrows in Figs.~\ref{Fig:6} ($a_1$) and ($a_4$), 
by observing that each density dip created in species A, Fig.~\ref{Fig:6} ($a_1$), is filled by a density 
hump (on top of the BEC background) developed in species B, Fig.~\ref{Fig:6} ($a_4$), and vice versa. 
Furthermore, it is found that consecutive orbitals within the same
species also follow the above-described phenomenology with a clearly visible domain-wall~\cite{darkbook, trippenbach} 
formed between the second and the 
third orbital of species B [see arrows in green in Figs.~\ref{Fig:6} ($a_4$)-($a_6$)]. 
For intermediate times the merging of the most inner solitary 
states discussed above is indicated with circles in Figs.~\ref{Fig:6} ($a_2$), ($a_5$). 
Notice the pronounced density hump that occurs in species B around the center of the trap, 
being supported by all three orbitals developed in this species.  
Additionally, also the faster DAD  
solitary waves are monitored in this time slice, 
where again it is observed that these states are 
supported by all orbitals used in each of the two species being    
marked with dashed rectangles.
However, at larger propagation times and since we ``kicked'' the system towards miscibility, 
multiple interference events more pronounced in species B, 
result to a dephasing of these matter wave patterns and most of these states are lost as can be seen 
in Figs.~\ref{Fig:6} ($a_3$), ($a_6$), 
rendering the two species mostly overlapped. 
Notice the increasing tendency towards miscibility with the overlap integral 
[see again here Fig.~\ref{Fig:6} ($b$) for $g_{AB}=0.5$] reaching its maximum 
value, $\Lambda_{MB}(t \geq 60) \approx 0.95$, at large propagation times, 
when compared to the MF approximation. 
In the latter case, $\Lambda_{MF}(t) \approx 0.85$ is reached 
from the early stages of the dynamics remaining on average almost the same
as time progresses. 

To conclude our investigation, let us also briefly comment on the manifestation of the 
MB correlated character of the quench-induced dynamics with the aid of in-situ single-shot measurements.
Figs.~\ref{Fig:7} $(a)$, $(b)$ present the first and the second simulated in-situ single-shot images at
$t_{im}=15$ for both species, with the DAD structures being clearly imprinted in both shots. Notice that the two 
species are almost completely overlapped resembling the overall tendency observed in the averaged, over $N_{shots}=1000$, 
one-body density illustrated in Fig. \ref{Fig:7} $(c)$.  
By inspecting the corresponding variances [see also Eqs. (\ref{Eq:7}) and (\ref{Eq:8})]
during the evolution shown in Fig. \ref{Fig:7} $(d)$, we observe that within the MF $\mathcal{V}_{MF}^A(t)$ and 
$\mathcal{V}_{MF}^B(t)$ exhibit a small 
amplitude oscillatory behavior reflecting the global breathing motion of each cloud.
Interestingly enough the oscillation amplitudes of $\mathcal{V}_{MF}^A(t)$ and $\mathcal{V}_{MF}^B(t)$
differ further, due to the difference in the magnitude of the breathing that each species undergoes 
[see also Figs. \ref{Fig:5} ($b_1$), ($b_2$)]. 
In sharp contrast to the above, the variances within the MB approach differ drastically from their MF counterparts. 
Indeed, both $\mathcal{V}_{MB}^A(t)$ and $\mathcal{V}_{MB}^B(t)$ show an overall increasing tendency 
indicating, as in the positive quench scenario, the presence of entanglement [see also the corresponding discussion 
in Sec. III B].   
Remarkably enough, $\mathcal{V}_{MB}^A(t)$ and $\mathcal{V}_{MB}^B(t)$ deviate significantly as 
a result of the strong intraspecies correlations.  
We should bear in mind that the initial pre-quenched state is both strongly fragmented and entangled on the MB level. 
Therefore, in this strongly correlated scenario both fragmentation as well as entanglement are greatly manifested in the 
evolution of the variance of a set of single-shot measurements. 

\section{Conclusions}\label{conclusions}
In the present work we explored the quench-induced phase separation dynamics of
an inhomogeneous repulsively interacting binary BEC both within and beyond the MF approximation 
including multiple orbitals. 
To achieve such a miscible to immiscible transition (positive quench case)
the intraspecies interactions are held fixed and the system is abruptly driven out-of-equilibrium
by switching on the interspecies repulsion.
Quench dynamics leads to the filamentation of the density of each of the two species
and also in both approaches (MF and MB) 
while the filaments formed perform collective oscillations of the
breathing-type. 
The wavenumbers associated with the observed growth
are identified to be shorter in the
MB case for all $g_{AB}$ values that we have checked, 
whilst our numerical findings at the MF level are in very good agreement with the 
analytical predictions available in this limit, as regards the instability
growth rate. 
It is found that increasing the interspecies repulsion, not only accelerates the filamentation 
process but also increases the number of filaments formed in both approaches, 
occurring faster on the MB level. 
Additionally, stronger interspecies repulsion leads to almost complete phase separation being 
more pronounced in the MB scenario. 
We further note, that upon fixing the interspecies repulsion while decreasing 
significantly the system size (few boson case) 
phase separation is absent in the MB case while 
still present at the MF limit.

Detailed correlation analysis at the one- and the two-body level
bear the signature of the phase separation process as the miscibility-immiscibility threshold is crossed.
On the one-body level significant losses of coherence are observed, verifying the fragmented nature of the system, 
between filaments residing around the center of the 
trap with the longer distant ones lying at the periphery of the bosonic cloud.
At the two-body level domain-wall-like structures are revealed, since the inner filaments in both species are 
found to be anti-correlated with their respective outer ones. These domain-walls
support the fact that for smaller interspecies interactions, but well inside the immiscible regime,
we never observe perfect de-mixing of the two species. 
Furthermore, and even more importantly, 
the presence of both entanglement and fragmentation are 
related to the variance of single-shot 
images, that are utilized for the first time in the current effort 
for binary systems, offering a direct way 
for the experimental realization of the observed dynamics.
In particular, it is found that the growth rate of the variance resembles 
the growth rate of the entanglement.  
The fragmentation of the binary system is captured by the deviations  
in the variance measured in the course of the dynamics 
with respect to each of the two species. 

Interestingly enough, when considering the reverse (negative) quench scenario, 
namely quenching from the immiscible towards the miscible 
regime multiple dark-antidark solitary waves are spontaneously generated in both approaches and they are found to decay in the 
MB case~\cite{lgs}. The evolution of the variance of single-shot measurements reveals  
enhanced entanglement, since the system in this case is strongly correlated on the MB level. 
Finally, for transitions inside the immiscible regime we retrieve the 1D analogue of the so-called ``ball''
and ``shell'' structure that appears in
higher-dimensional binary BECs~\cite{usadsh,Bisset}. 

There are multiple directions that are of interest for future work along the lines of the current effort. 
A systematic study of the dynamical phase separation process following a time-dependent protocol 
(e.g. a linear quench) presents one of the major computational challenges for further study. In particular, in such a scenario
one can explore the domain formation crossing the critical point with different velocities and thus testing the Kibble-Zurek
mechanism~\cite{Sabbatini} in the presence of quantum fluctuations.
However, to examine the latter, a major challenge that it is imperative
to overcome is that of considering low atom numbers, in order
to explore the associated thermodynamic limit, avoiding the potential
influence of finite size effects.
Another straight forward direction is to consider the corresponding already experimentally 
realized~\cite{wieman} 2D setting, and examine how the MF properties are altered in the presence of quantum fluctuations. 
Also of great interest would be to consider the quench dynamics of spinor 
BECs, for which phase separation processes are of ongoing interest at the MF limit~\cite{gautam} 
and also investigate the relevant MB aspects.

\appendix

\section{Single-Shot Measurements in Binary Bosonic Mixtures} \label{sec:singleshots} 

As in the single component case, the single-shot simulation procedure relies on a sampling of the 
MB probability distribution \cite{kaspar,Lode,filled_vortex}.  
The latter is available within the ML-MCTDHB framework. 
However, in a two-species BEC and when inter and intraspecies correlations are taken into account, 
the entire single-shot procedure is significantly altered when compared to the single component case.   
Here, the role of entanglement between the species manifested by the Schmidt decomposition 
[see Eq. (\ref{Eq:3})] and in particular the Schmidt coefficients $\lambda_k$'s play a 
crucial role concerning the image ordering. 
\begin{figure*}[ht]
\includegraphics[width=0.86\textwidth]{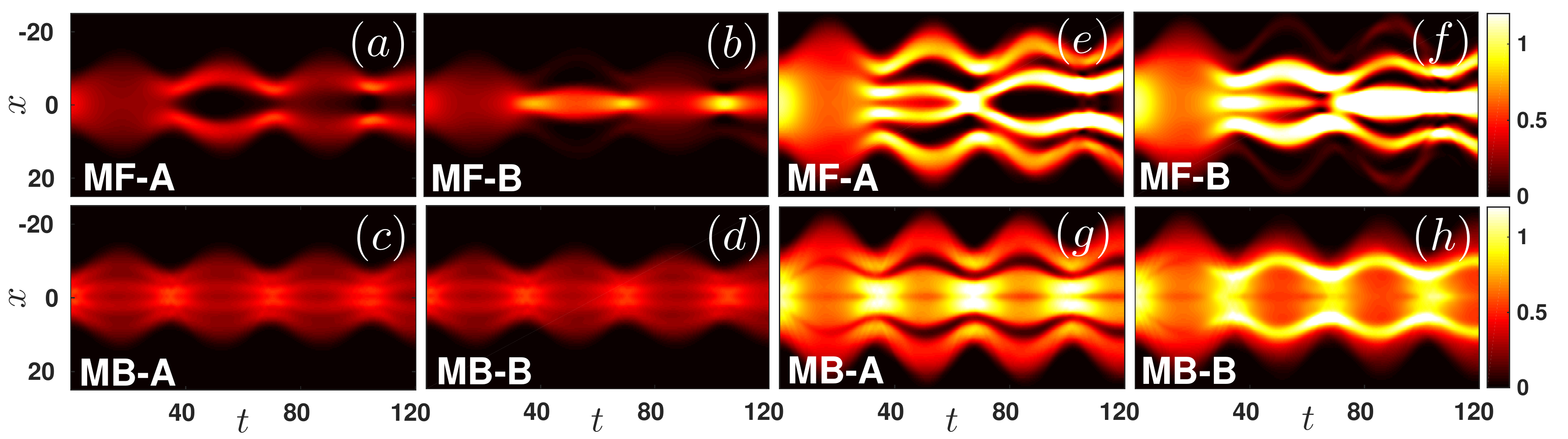}
\caption{ %(Color online) 
($a$), ($b$) [($c$), ($d$)]
Quenched $\rho^{(1),A}(x;t)$, and $\rho^{(1),B}(x;t)$ from the miscible ($g_{AB}=0$) 
to the immiscible phase ($g_{AB}=1.2$) 
obtained within the MF [MB] approach 
for $N_A=N_B=5$ atoms.  
($e$), ($f$) [($g$), ($h$)] The same as the above but for $N_A=N_B=20$ atoms.
Other parameters used are the same as in Fig.~\ref{Fig:1}.} 
\label{Fig:8}
\end{figure*}

For instance, to image first the $A$ and then the $B$ species we consecutively annihilate all 
the $N_A$ particles. 
Focusing first on a certain imaging time instant, $t_{im}$, 
a random position is drawn according to the constraint $\rho_{N_A}^{(1)}(x_1')>l_1$ where $l_1$ 
refers to a random number within the interval 
[$0$, $ \max\lbrace{\rho^{(1)}_{N_A}(x;t_{im})\rbrace}$]. 
Then we project the ($N_A+N_B$)-body wavefunction to the ($N_A-1+N_B$)-body one, by employing the 
operator $\frac{1}{\mathcal{N}}(\hat{\Psi}_A(x_1')\otimes \hat{\mathbb{I}}_B)$, where 
$\hat{\Psi}_A(x_1')$ denotes the bosonic field operator that annihilates an $A$ species boson at position $x_1'$ and 
$\mathcal{N}$ is the normalization constant.  
The latter process directly affects the 
$\lambda_k$'s (entanglement weights) and thus 
despite the fact that the $B$ species has not been imaged yet, both $\rho^{(1)}_{N_A-1}(t_{im})$ and $\rho^{(1)}_{N_B}
(t_{im})$ change. 
This can be easily understood by employing once more the Schmidt decomposition. 
Indeed after this first measurement the MB wavefunction reads
\begin{equation}
\begin{split}
&\ket{\tilde{\Psi}_{MB}^{N_A-1,N_B}(t_{im})}=\\ &\sum_i \sqrt{\tilde{\lambda}_{i,N_A-1}(t_{im})}\ket{\tilde{\Psi}_{i,N_A-1}^A(t_{im})}\ket{\Psi_i^B(t_{im})},  
\label{Eq:A1}
\end{split}
\end{equation}
where $\ket{\tilde{\Psi}_{i,N_A-1}^A}=\frac{1}{N_i}\hat{\Psi}_A(x_1')\ket{\Psi_i^A}$ is the 
$N_A-1$ species wavefunction. 
$N_i=\sqrt{\bra{\Psi_i^A}\hat{\Psi}_A^{\dagger}(x_1')\hat{\Psi}_A(x_1')\ket{\Psi_i^A}}$ denotes the normalization factor and 
$\tilde{\lambda}_{i,N_A-1}=\lambda_i N_i/\sum_i \lambda_i N_i^2$ are the Schmidt coefficients that refer to the 
($N_A-1+N_B$)-body wavefunction. 
The above-mentioned procedure is repeated for $N_A-1$ steps and the resulting distribution of 
positions ($x'_1$, $x'_2$,...,$x'_{N_A-1}$) is convoluted with a point spread function 
leading to a single-shot $\mathcal{A}^A(\tilde{x})=\sum_{i=1}^{N_A}e^{-\frac{(\tilde{x}-x'_i)^2}{2w^2}}$ for the $A$ species. 
Here $\tilde{x}$ refers to the spatial coordinates within the image and $w$ is the width of the point spread function. 
It is worth mentioning also at this point that before annihilating the last of the $N_A$ 
particles, the MB wavefunction has the form 
\begin{equation}
\begin{split}
\ket{\tilde{\Psi}_{MB}^{1,N_B}(t_{im})}=\sum_i \sqrt{\tilde{\lambda}_{i,1}
(t_{im})}\ket{\Phi_{i,1}^A(t_{im})}\ket{\Psi_i^B(t_{im})},   
\label{Eq:A2}
\end{split}
\end{equation}
where $\ket{\Phi_{i,1}^A(t_{im})}$ denotes a single particle wavefunction characterizing the $A$ species.  
Then, it can be easily shown that annihilating the last $A$ species particle the MB wavefunction reads 
\begin{equation}
\begin{split}
&\ket{\tilde{\Psi}_{MB}^{0,N_B}(t_{im})}=\\ &\ket{0} \otimes\sum_i \frac{\sqrt{\tilde{\lambda}_{i,1}(t_{im})}
\braket{x|\Phi_{i,1}^A}}{\sum_j{\sqrt{\tilde{\lambda}_{j,1}(t_{im})|\braket{x|\Phi_{j,1}^A}|^2}}}\ket{\Psi_i^B(t_{im})},   
\label{Eq:A3}
\end{split}
\end{equation}
where $\braket{x|\Phi_{j,1}^A}$ is the single particle orbital of the $j$-th mode.  
After this last step the entanglement between the species has been destroyed and the wavefunction of 
the B species $\ket{\Psi_{MB}^{N_B}(t_{im})}$ corresponds to the second term of the cross product on the right hand side of Eq. (\ref{Eq:A3}).   
In this way, it becomes evident that $\ket{\Psi_{MB}^{N_B}(t_{im})}$ obtained after the annihilation 
of all $N_A$ atoms is a non-entangled $N_B$-particle MB wavefunction and its corresponding single-shot procedure is 
the same as in the single species case \cite{kaspar}.  
The latter is well-established (for details see \cite{kaspar,Lode}) and therefore it is only 
briefly outlined below. 
Referring to $t=t_{im}$ we first calculate $\rho^{(1)}_{N_B}(x;t_{im})$ 
from the MB wavefunction $\ket{\Psi_{N_B}}\equiv \ket{\Psi(t_{im})}$.  
Then, a random position $x''_1$ is drawn obeying $\rho^{(1)}_{N_B}(x''_1;t_{im})>l_2$ where $l_2$ is    
a random number in the interval [$0$, $\rho^{(1)}_{N_B}(x;t_{im})$].   
Next, one particle located at a position $x''_1$ is annihilated and $\rho^{(1)}_{N_B-1}(x;t_{im})$  
is calculated from $\ket{\Psi_{N_B-1}}$.  
To proceed, a new random position $x''_2$ is drawn from $\rho^{(1)}_{N_B-1}(x;t_{im})$.  
Following this procedure for $N_B-1$ steps we obtain the distribution of positions 
($x''_1$, $x''_2$,...,$x''_{N_B-1}$) which is then convolved with a point spread function  
resulting in a single-shot $\mathcal{A}^B(\tilde{x'}|\mathcal{A}^A(\tilde{x}))$.  

We remark here that the same overall procedure can be followed in order first to image the 
$B$ and then the $A$ species. 
Such an imaging process results in the corresponding single-shots $\mathcal{A}^B(\tilde{x})$ and 
$\mathcal{A}^A(\tilde{x'}|\mathcal{A}^B(\tilde{x}))$. 

\section{Few boson case} \label{few}
Here, we explore the dependence of a miscible-immiscible transition, from $g_{AB}=0$
to $g_{AB}=1.2$, on the total
number of atoms, $N$, of the binary system.  
Initially we consider a binary system consisting of $N=40$ atoms, 
which is almost half the total number of particles considered in the main text
($N=100$), and as a next step  
a mixture with $N=10$ bosons, i.e. an order of magnitude smaller cloud, is studied.
Our findings are summarized in Fig.~\ref{Fig:8}.   
At the MF level depicted in Figs.~\ref{Fig:8} $(a),(b)$ and $(e),(f)$ for $N=10$ and $N=40$
respectively, 
we find that the number of filaments formed depends on the number of atoms present in the system
and for larger particle numbers more filaments are formed. 
In sharp contrast to the above dynamics, for small particle 
numbers, i.e. $N=10$, 
phase separation is not observed in the MB approach
(while it is transparent at the MF level in the form of
a ball and shell configuration);
instead an enhanced miscibility region
is evident in Figs.~\ref{Fig:8} $(c),(d)$. 
Alterations of the miscibility-immiscibility threshold due to the presence of quantum pressure effects in confined BECs  
have been reported in~\cite{navarro,Proukakis,wen} but at the MF level.
Remarkably here, and also in contrast to the MF approximation
four, instead of two, almost equally populated filaments are dynamically formed in both the A and the B species shown 
respectively in Figs.~\ref{Fig:8} $(c)$, and $(d)$, but  
the two species remain overlapping at all times. 
Additionally, the interparticle repulsion between the species 
leads to breathing-type oscillations of the particle densities. 
\begin{figure*}[ht]
\includegraphics[width=0.9\textwidth]{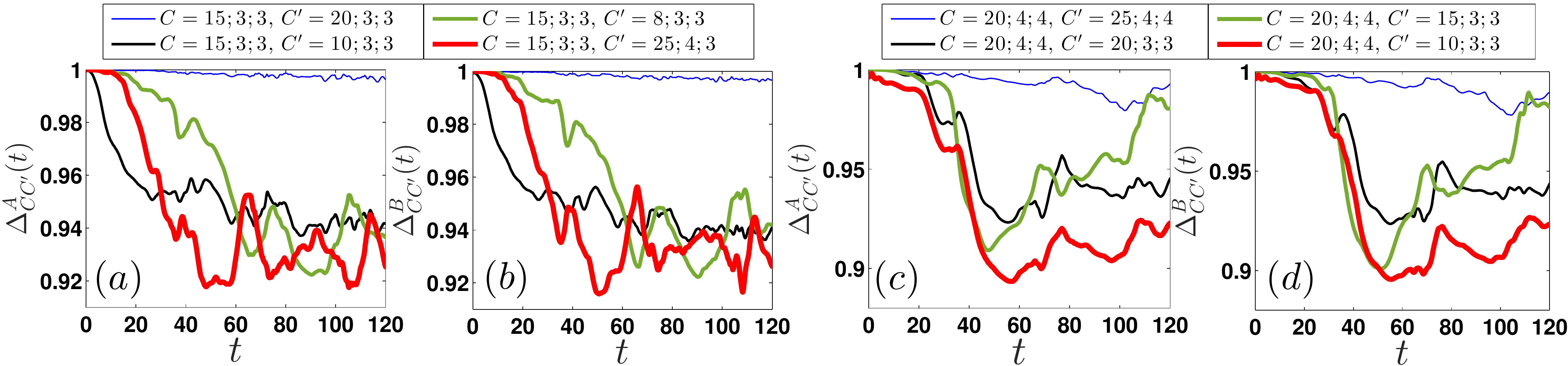}
\caption{ %(Color online) 
Evolution of the relative overlap of the one-body densities $\Delta_{CC'}^{\sigma}(t)$ between 
different numerical configurations $C$ and $C'$ (see legends) for the ($a$) $\sigma=A$ and ($b$) $\sigma=B$ species in the 
case of $N_A=N_B=50$ bosons.  
($c$), ($d$) The same as ($a$) and ($b$) but for $N_A=N_B=20$. } 
\label{Fig:9}
\end{figure*}

As the number of particles is increased, namely for $N=40$,  
the one-body density evolution of the $A$ species shown in Figs.~\ref{Fig:8} $(e)$, $(g)$ for the MF and the MB scenario
respectively also differ. In particular, while in both approaches four filaments are formed, they are found to be 
significantly broader in the MB case.
This broadening together with the breathing that the cloud undergoes, 
leads to an attraction, collision, and repulsion of the 
inner filaments in a periodic manner, 
being more pronounced in the MB case when compared to the single merging, and repulsion observed at 
around $t\approx 70$ in the MF approach of Fig.~\ref{Fig:8} $(e)$. 
Moreover, the disparity between the two approaches becomes rather transparent when further inspecting the spatio-temporal 
evolution of the density of species $B$ illustrated in Figs.~\ref{Fig:8} $(f)$, $(h)$ for the MF and the MB case 
respectively. Interestingly here, in the MB scenario
only two filaments are formed located alternately in regions 
that correspond to density dips of species $A$, 
restoring the phase separation process absent for smaller particle numbers. 
However, the central filament created in the MF approach 
[see for comparison Fig.~\ref{Fig:8} $(f)$] is clearly  
absent in the MB case, resulting in this way in a larger overlap between the two gases at the MB level.

\section{Remarks on Convergence} \label{sec:numerics1}

Let us first briefly comment on the main features of our computational methodology, ML-MCTDHB, 
and then showcase the convergence of our results. 
ML-MCTDHB \cite{Lushuai,MLX} constitutes a flexible variational method 
for solving the time-dependent MB Schr{\"o}dinger equation of bosonic mixtures. 
It relies on expanding the total MB wavefunction with respect to a 
time-dependent and variationally optimized basis, which enables us to capture the 
important correlation effects using a computationally feasible basis size. 
Finally, its multi-layer ansatz for the total wavefunction allows us to account 
for intra- and interspecies correlations when simulating the dynamics of bipartite systems. 
For our simulations, we use a primitive basis consisting of 
a sine discrete variable representation containing 800 grid points.  
To perform the simulations into a finite spatial region, we 
impose hard-wall boundary conditions at the positions $x=\pm50$.
Note that the Thomas-Fermi radius of each bosonic cloud is of the order of 20
and we never observe appreciable densities beyond $x=\pm 30$.
Therefore the location of the imposed boundary conditions
is inconsequential for our simulations.
The truncation of the total system's Hilbert space, namely the order of the considered 
approximation, is indicated by the used numerical configuration space $C=(M;m_A;m_B)$.  
Here, $M=M_A=M_B$ refers to the number of species functions and $m_A$, $m_B$ denote 
the amount of SPFs for each of the species. 
In the limit $M=m_A=m_B=1$ the ML-MCTDHB expansion reduces to the MF ansatz. 
Finally, in order to guarantee the accurate performance of 
the numerical integration for the ML-MCTDHB equations of motion the following overlap 
criteria $|\langle \Psi |\Psi \rangle -1| < 10^{-10}$ and
$|\langle \varphi_i |\varphi_j \rangle -\delta_{ij}| < 10^{-10}$ have been imposed for the 
total wavefunction and the SPFs respectively. 

Next, we demonstrate the order of convergence of our results 
and thus the level of our MB truncation scheme. 
To show that our MB results (more specifically the quantities and observables considered here)
are numerically converged, we inspect for the $\sigma$ species the 
overlap $\Delta_{CC'}^{\sigma}(t)=1-\delta_{CC'}^{\sigma}(t)$ between the one-body densities $\rho_i^{(1),\sigma}(x,t)$, where $i=C, C^{'}$,   
obtained within the different numerical configurations $C=(M;m_A;m_B)$ and $C'=(M';m_A';m_B')$ 
\begin{equation}
\delta_{CC'}^{\sigma}(t)=\frac{1}{N_{\sigma}}\int_R dx \left(\rho_C^{(1),\sigma}(x,t)-\rho_{C'}^{(1),\sigma}(x,t)\right).
\label{Eq:B1} 
\end{equation}
$N_{\sigma}$ denotes the number of $\sigma$ species bosons and $R=[-30,30]$ corresponds to the 
spatially integrated domain in which there is finite density.
In this way, we track the relative error between the different approximations $C$, $C'$ and infer about 
convergence when $\Delta_{CC'}^{\sigma}(t)$ becomes to a certain degree insensitive upon increasing 
either the number of species functions $M$ or the SPFs $m_A$, $m_B$. 
$\Delta_{CC'}^{\sigma}$ is bounded within the interval $[0,1]$, where  
in the case of $\Delta_{CC'}^{\sigma}=1$ [$\Delta_{CC'}^{\sigma}=0$] the two densities completely 
overlap [phase separate] and therefore the $C$, $C'$ approximations yield the same [deviating] results. 
Figs. \ref{Fig:9} ($a$), ($b$) present $\Delta_{CC'}^A(t)$ and $\Delta_{CC'}^B(t)$ respectively, for $N_A=N_B=50$ and 
post-quench interspecies interaction $g_{AB}=1.2$. 
Here, we keep always $C=(15;3;3)$ fixed and examine the convergence upon varying either $M'$ or $m_A'$, $m_B'$.  
As it can be seen, upon increasing the number of species functions from $M=15$ to $M=20$, i.e. 
$C=(15;3;3)$ and $C'=(20;3;3)$, $\Delta_{CC'}^A(t)$ [$\Delta_{CC'}^B(t)$] exhibits negligible deviations 
being smaller than $1\%$ throughout the dynamics. 
Therefore convergence is guaranteed with respect to $M$. 
However, for increasing number of SPFs $\Delta_{CC'}^{\sigma}(t)$ is more sensitive. 
Indeed, by considering $C'=(25;4;3)$ corresponding to a total number of coefficients 625025  
[instead of 44805 that refer to the $C=(15;3;3)$] the deviation obtained from 
$\Delta_{CC'}^A(t)$ [$\Delta_{CC'}^B(t)$] reaches a maximum value of the order of $8\%$ at large propagation times. 
We should note here that further increase of the number of SPFs is computationally prohibitive 
for this number of particles as the considered number of configurations becomes significantly larger.  
The same observations can also be obtained from $\Delta_{CC'}^{\sigma}(t)$ of a mixture consisting 
of $N_A=N_B=20$ bosons, see Figs. \ref{Fig:9} ($c$), ($d$), when considering $C=(20;4;4)$.  
For completeness we note that fragmentation becomes enhanced all the more as
the particle number is reduced.  
To conclude upon convergence concerning the species functions we show $\Delta_{CC'}^A(t)$ [$\Delta_{CC'}^B(t)$] 
in Fig. \ref{Fig:9} ($c$) [($d$)].  
It is observed that $\Delta_{CC'}^A(t)$ [$\Delta_{CC'}^A(t)$] between $C=(20;4;4)$ and $C'=(25;4;4)$ testifies 
negligible deviations which become at most $2.2\%$ at long evolution times.  
In the same manner, convergence occurs for a varying number of SPFs in both species. 
For instance, $\Delta_{CC'}^A(t)$ [$\Delta_{CC'}^B(t)$] between $C'=(20;3;3)$ and $C=(20;4;4)$ shows 
a maximum deviation of the order of $7\%$ for large evolution times. 
Similar observations can be deduced also for the case of even smaller particle numbers,
and the reverse quench scenario (not included here for brevity reasons). 
To summarize, according to the above systematic investigations, the considered orbital configurations provide 
adequate approximations for the description of the non-equilibrium correlated dynamics.

\section*{Acknowledgements} 
S.I.M. and P.S. gratefully acknowledge financial support by the Deutsche Forschungsgemeinschaft 
(DFG) in the framework of the SFB 925 ``Light induced dynamics and control of correlated quantum
systems''.  
P.G.K. gratefully acknowledges the support of NSF-PHY-1602994 and the
Alexander von Humboldt Foundation. 
S.I.M. and G.C.K. would like to thank G. M. Koutentakis for fruitful discussions.

{}


\begin{thebibliography}{60}


\bibitem{pethick}C. J. Pethick and H. S. Smith,
\emph{Bose-Einstein Condensation in Dilute Gases}, Cambridge
University Press, Cambridge, 2002.

\bibitem{stringari} L. P. Pitaevskii and S. Stringari,
{\it Bose-Einstein Condensation}, Oxford University Press (Oxford, 2003).


\bibitem{emergent}
P. G. Kevrekidis, D. J. Frantzeskakis, and R. Carretero-Gonz{\'a}lez (eds.),
{\it Emergent nonlinear phenomena in Bose-Einstein condensates. Theory and experiment}
(Springer-Verlag, Berlin, 2008).

\bibitem{darkbook} P. G. Kevrekidis,
D. J. Frantzeskakis, and R. Carretero-Gonz{\'a}lez,
{\it The Defocusing Nonlinear Schr{\"o}dinger Equation},
SIAM (Philadelphia, 2015).


\bibitem{lcarr} L. D. Carr, {\it Understanding Quantum Phase Transitions},
(Taylor \& Francis, Boca Raton, 2010).

\bibitem{prouka}  N. Proukakis, S. Gardiner, M. Davis, and M. Szymanska,
{\it Quantum gases: finite temperature and non-equilibrium dynamics},
(Imperial College Press, London, 2013).

\bibitem{ueda} Y. Kawaguchi, and M. Ueda,
  Phys. Rep. {\bf 250}, 253 (2012).

  \bibitem{nake} D. M. Stamper-Kurn, 
M. R. Andrews, A. P. Chikkatur, S. Inouye, H.-J. Miesner, J. Stenger, and 
W. Ketterle, Phys.\ Rev.\ Lett.\ \textbf{80}, 2027 (1998).



\bibitem{dsh} D. S.\ Hall, 
M. R. Matthews, J. R. Ensher, C. E. Wieman, and 
E. A. Cornell, \newblock Phys.\ Rev.\ Lett.\ 
\textbf{81}, 1539 (1998).

\bibitem{stenger} J. Stenger, S. Inouye, D. M. Stamper-Kurn,
H. -J. Miesner, A. P. Chikkatur, and W. Ketterle,
Nature {\bf 396}, 345 (1998).


\bibitem{cornell} V. Schweikhard, I. Coddington, P. Engels, S. Tung,
and E. A. Cornell, Phys. Rev. Lett. {\bf 93}, 210403 (2004).

\bibitem{usadsh} K. M. Mertes, J. W. Merrill, R. Carretero-Gonz{\'a}lez,
D. J. Frantzeskakis, P. G. Kevrekidis, and D. S. Hall,
Phys. Rev. Lett. {\bf 99}, 190402 (2007).

\bibitem{wieman}  S. B. Papp, J. M. Pino, and C. E. Wieman,
Phys. Rev. Lett. {\bf 101}, 040402 (2008).

\bibitem{hall2} R. P. Anderson, C. Ticknor, A. I. Sidorov,
and B. V. Hall, Phys. Rev. A {\bf 80}, 023603 (2009).

\bibitem{hall3} M. Egorov, B. Opanchuk, P. Drummond, B. V. Hall, P. Hannaford, and A. I. Sidorov,
Phys. Rev. A {\bf 87}, 053614 (2013).

\bibitem{tojo} S. Tojo, Y. Taguchi, Y. Masuyama, T. Hayashi,
H. Saito, and T. Hirano, Phys. Rev. A {\bf 82}, 033609 (2010).

\bibitem{eto1} Y. Eto, M. Takahashi, M. Kunimi, H. Saito, and T. Hirano, 
%Nonequilibrium dynamics induced by miscible–immiscible transition in binary Bose–Einstein condensates. 
New J. Phys. \textbf{18}, 073029 (2016). 

\bibitem{eto2} Y. Eto, M. Takahashi, K. Nabeta, R. Okada, M. Kunimi, H. Saito, and T. Hirano, 
%Bouncing motion and penetration dynamics in multicomponent Bose-Einstein condensates. 
Phys. Rev. A \textbf{93}, 033615 (2016). 


\bibitem{flop} E. Nicklas, H. Strobel, T. Zibold, C. Gross,
B. A. Malomed, P. G. Kevrekidis, M. K. Oberthaler,
Phys. Rev. Lett. {\bf 107}, 193001 (2011).

\bibitem{strobel} E. Nicklas, W. Muessel, H. Strobel, P. G. Kevrekidis,
and M. K. Oberthaler, Phys. Rev. A {\bf 92}, 053614 (2015).

\bibitem{spielman} Y. -J. Lin, K. Jim{\'e}nez-Garc{\'i}a,
I. B. Spielman, Nature (London) {\bf 471}, 83 (2011).

\bibitem{gas1} E. Nicklas, M. Karl, M. H{\"o}fer, A. Johnson, W. Muessel, H. Strobel, J. Tomkovic, 
T. Gasenzer, and M. K. Oberthaler,
Phys. Rev. Lett. {\bf 115}, 245301 (2015). 

\bibitem{gas2} M. Karl, H. Cakir, J. C. Halimeh, M. K. Oberthaler, M. Kastner, and T. Gasenzer,
Phys. Rev. E {\bf 96}, 022110 (2017).

\bibitem{bisset} R. N. Bisset, R. M. Wilson, and C. Ticknor,
Phys. Rev. A {\bf 91}, 053613 (2015).


\bibitem{March} M. A. Garc{\'i}a-March, B. Juli{\'a}-D{\'i}az, G. E. Astrakharchik, J. Boronat, and A. Polls, 
%Distinguishability, degeneracy, and correlations in three harmonically trapped bosons in one dimension. 
Phys. Rev. A \textbf{90}, 063605 (2014). 


\bibitem{Cazalilla} M. A. Cazalilla, and A. F. Ho, 
%Instabilities in binary mixtures of one-dimensional quantum degenerate gases. 
Phys. Rev. Lett. \textbf{91}, 150403 (2003). 


\bibitem{Alon_mixt} O. E. Alon, A. I. Streltsov, and L. S. Cederbaum, 
%Demixing of bosonic mixtures in optical lattices from macroscopic to microscopic scales. 
Phys. Rev. Lett. \textbf{97}, 230403 (2006). 


\bibitem{March1} M. A. Garc{\'i}a-March, and T. Busch, 
%Quantum gas mixtures in different correlation regimes. 
Phys. Rev. A \textbf{87}, 063633 (2013). 


\bibitem{Mishra} T. Mishra, R. V. Pai, and B. P. Das, 
%Phase separation in a two-species Bose mixture. 
Phys. Rev. A \textbf{76}, 013604 (2007). 


\bibitem{comp_ferm} S. Z{\"o}llner, H. D. Meyer, and P. Schmelcher, 
%Composite fermionization of one-dimensional Bose-Bose mixtures. 
Phys. Rev. A \textbf{78}, 013629 (2008). 

\bibitem{Pyzh} M. Pyzh, S. Kr{\"o}nke, C. Weitenberg, and P. Schmelcher, 
%Spectral properties and breathing dynamics of a few-body Bose-Bose mixture in a 1D harmonic trap. arXiv preprint 
New J. Phys. {\bf 20}, 015006 (2018).  


\bibitem{Hao} Y. Hao, and S. Chen, 
%Density-functional theory of two-component Bose gases in one-dimensional harmonic traps. 
Phys. Rev. A \textbf{80}, 043608 (2009). 


\bibitem{March2} M. A. Garc{\'i}a-March, B. Juli{\'a}-D{\'i}az, G. E. Astrakharchik, T. Busch, J. Boronat, and A. Polls, 
%Sharp crossover from composite fermionization to phase separation in microscopic mixtures of ultracold bosons. 
Phys. Rev. A \textbf{88}, 063604 (2013). 


\bibitem{March3} M. A. Garc{\'i}a-March, B. Juli{\'a}-D{\'i}az, G. E. Astrakharchik, T. Busch, J. Boronat, and A. Polls, 
%Quantum correlations and spatial localization in one-dimensional ultracold bosonic mixtures. 
New J. Phys. \textbf{16}, 103004 (2014). 


\bibitem{Pflanzer} A. C. Pflanzer, S. Z{\"o}llner, and P. Schmelcher, 
%Material-barrier tunnelling in one-dimensional few-boson mixtures. 
J. Phys. B: At.  Mol. Opt. Phys. \textbf{42}, 231002 (2009). 



\bibitem{Pflanzer1} A. C. Pflanzer, S. Z{\"o}llner, and P. Schmelcher, 
%Interspecies tunneling in one-dimensional Bose mixtures. 
Phys. Rev. A \textbf{81}, 023612 (2010). 


\bibitem{Chatterjee} B. Chatterjee, I. Brouzos, L. Cao, and P. Schmelcher, 
%Few-boson tunneling dynamics of strongly correlated binary mixtures in a double well. 
Phys. Rev. A \textbf{85}, 013611 (2012). 

\bibitem{Campbell} S. Campbell, M. {\'A}. Garc{\'i}a-March, T. Fogarty, and T. Busch, 
%Quenching small quantum gases: Genesis of the orthogonality catastrophe. 
Phys. Rev. A \textbf{90}, 013617 (2014). 


\bibitem{Mistakidis} S.I. Mistakidis, L. Cao, and P. Schmelcher, 
%Interaction quench induced multimode dynamics of finite atomic ensembles. 
J. Phys. B: At., Mol. Opt. Phys. \textbf{47}, 225303 (2014). 


\bibitem{Mistakidis1} S.I. Mistakidis, L. Cao, and P. Schmelcher, 
%Negative-quench-induced excitation dynamics for ultracold bosons in one-dimensional lattices. 
Phys. Rev. A \textbf{91}, 033611 (2015). 


\bibitem{Mistakidis5} S.I. Mistakidis, T. Wulf, A. Negretti, and P. Schmelcher, 
%Resonant quantum dynamics of few ultracold bosons in periodically driven finite lattices. 
J. Phys. B: At. Mol. Opt. Phys. \textbf{48}, 244004 (2015). 


\bibitem{Mistakidis2} S.I. Mistakidis, and P. Schmelcher, 
%Mode coupling of interaction quenched ultracold few-boson ensembles in periodically driven lattices. 
Phys. Rev. A \textbf{95}, 013625 (2017). 

\bibitem{Lushuai}
L. Cao, S. Kr{\" o}nke, O. Vendrell, and P. Schmelcher,
%{\it The multi-layer multi-configuration time-dependent Hartree method for bosons: Theory, implementation, and applications.}
J. Chem. Phys. \textbf{139}, 134103  (2013).

\bibitem{MLX}  L. Cao, V. Bolsinger, S. I. Mistakidis, G. M. Koutentakis, S. Kr{\"o}nke, J. M. Schurer, and P. Schmelcher,
%A unified ab-initio approach to the correlated quantum dynamics of ultracold fermionic and bosonic mixtures 
J. Chem. Phys. \textbf{147}, 044106 (2017). 

\bibitem{trippenbach} M. Trippenbach, K. G{\'o}ral, K. Rz\c{a}$\rm{\dot{z}}$ewski,
B. A. Malomed, and Y. B. Band,
J. Phys. B: At. Mol. Opt. Phys. {\bf 33}, 4017 (2000).


\bibitem{boris} G. Filatrella, B.A. Malomed, M. Salerno,
  Phys. Rev. A {\bf 90}, 043629 (2014).

\bibitem{de}  S. De, D.L. Campbell, R.M. Price,
  A. Putra, B.M. Anderson, I.B. Spielman,
  Phys. Rev. A {\bf 89}, 033631 (2014).

\bibitem{Danaila} I. Danaila, M. A. Khamehchi, V. Gokhroo, P. Engels, and P. G. Kevrekidis, 
%Vector dark-antidark solitary waves in multicomponent Bose-Einstein condensates. 
Phys. Rev. A \textbf{94}, 053617 (2016). 


\bibitem{KevrekidisDAD} P. G. Kevrekidis, H. E. Nistazakis, D. J. Frantzeskakis, B. A. Malomed, and R. Carretero-González, 
%Families of matter-waves in two-component Bose-Einstein condensates. 
Europ. Phys. J. D-At., Mol., Opt. and P. Phys., \textbf{28}, 181 (2004). 


\bibitem{lgs} G. C. Katsimiga, G. M. Koutentakis, S. I. Mistakidis, P. G. Kevrekidis,
and P. Schmelcher, New J. Phys. {\bf 19}, 073004 (2017).
%Dark–bright soliton dynamics beyond the mean-field approximation


\bibitem{Kohler} T. K{\"o}hler, K. Goral, and P. S. Julienne, Rev. Mod. Phys. \textbf{78}, 1311 (2006). 

\bibitem{Chin} C. Chin, R. Grimm, P. Julienne, and E. Tiesinga, Rev. Mod. Phys. \textbf{82}, 1225 (2010).

\bibitem{Olshanii} M. Olshanii, Phys. Rev. Lett. \textbf{81}, 938 (1998). 

\bibitem{Kim} J. I. Kim, V. S. Melezhik, and P. Schmelcher, Phys. Rev. Lett. \textbf{97}, 193203 (2006). 

\bibitem{Frenkel} J. Frenkel, {\it in Wave Mechanics} 1st ed. (Clarendon Press, Oxford, 1934), pp. 423-428.

\bibitem{Dirac} P. A. Dirac,  
Proc. Camb. Phil. Soc., 
\textbf{26}, 376, Cambridge University Press (1930). 

 
\bibitem{Horodecki} R. Horodecki, P. Horodecki, M. Horodecki, and K. Horodecki,  
%Quantum entanglement. 
Rev. Mod. Phys. \textbf{81}, 865 (2009). 

\bibitem{note_ent} Commonly used measures to quantify bipartite entanglement are the von-Neumann 
entropy, $S[\Psi_{MB}(t)]=-\sum_k\lambda_k(t)\log(\lambda_k(t))$, 
and the concurence $D[\Psi_{MB}(t)]=2\sum_{i<j}\sqrt{\lambda_i(t) \lambda_j(t)}$. 
Here, $\rho^{\sigma}(t)=\sum_k\lambda_k(t)\ket{\Psi_k^{\sigma}(t)}\bra{\Psi_k^{\sigma}(t)}$ refers to the $N$-body density matrix and 
$\Psi_k^{\sigma}(t)$ denotes the $k$-th species function of the $\sigma$ species. Note that both measures  
vanish in the non entangled case. 

\bibitem{Roncaglia} M. Roncaglia, A. Montorsi, and M. Genovese, Phys. Rev. A \textbf{90}, 062303 (2014). 

\bibitem{Peres} A. Peres, {\it Quantum theory: concepts and methods} (Vol. 57), Springer Science and Business Media (2006). 

\bibitem{Lewenstein1} M. Lewenstein, D. Bru$\ss$, J. I. Cirac, B. Kraus, M. Ku$s'$, J. Samsonowicz, A. Sanpera, and R. 
Tarrach, 
%Separability and distillability in composite quantum systems-a primer. 
J. Mod. Opt. \textbf{47}, 2481 (2000). 


\bibitem{note1} 
The general ML-MCTDHB ansatz for a bosonic mixture consisting of an arbitrary number of component has 
been introduced in Ref. \cite{Lushuai}. 
Here, we utilize the Schmidt decomposition that holds for binary mixtures.  


\bibitem{mueller} E. J. Mueller, T. L. Ho, M. Ueda, and G. Baym, 
Phys. Rev. A \textbf{74}, 033612 (2006).


\bibitem{penrose} O. Penrose, and L. Onsager, Phys. Rev. \textbf{104}, 576 (1956).

\bibitem{aochui} P. Ao and S. T. Chui
Phys. Rev. A {\bf 58}, 4836 (1998).

\bibitem{navarro} R. Navarro, R. Carretero-Gonz{\'a}lez, and P. G. Kevrekidis,
Phys. Rev. A {\bf 80}, 023613 (2009).

 \bibitem{Abraham} J.W. Abraham, and M. Bonitz, 
%Quantum Breathing Mode of Trapped Particles: From Nanoplasmas to Ultracold Gases. 
Contrib. Plasm. Phys. \textbf{54}, 27 (2014). 

	
\bibitem{Tommasini} P. Tommasini, E. J. V. de Passos, A. F. R. de Toledo Piza, M. S. Hussein, and E. Timmermans,
Phys. Rev. A {\bf 67}, 023606 (2003). 
%Bogoliubov theory for mutually coherent condensates.

\bibitem{Sabbatini} J. Sabbatini, W. H. Zurek, and M. J. Davis,
Phys. Rev. Lett. {\bf 107}, 230402 (2011).
%Phase Separation and Pattern Formation in a Binary Bose-Einstein Condensate.
 
\bibitem{Timmermans} E. Timmermans,
Phys. Rev. Lett. {\bf 81}, 5718 (1998). 
%Phase Separation of Bose-Einstein Condensates.


\bibitem{Grond_lr} J. Grond, A. I. Streltsov, A. U. Lode, K. Sakmann, L. S. Cederbaum, and O. E. Alon, 
%Excitation spectra of many-body systems by linear response: General theory and applications to trapped condensates. 
Phys. Rev. A \textbf{88}, 023606 (2013).  


\bibitem{jain} P. Jain and M. Boninsegni,
Phys. Rev. A {\bf 83}, 023602 (2011).
%Quantum de-mixing in binary mixtures of dipolar bosons

\bibitem{Bandyopadhyay} S. Bandyopadhyay, A. Roy, and D. Angom,
Phys. Rev. A {\bf 96}, 043603, (2017).
% Dynamics of phase separation in two species Bose-Einstein condensates with vortices

\bibitem{kaspar} K. Sakmann, and M. Kasevich,
%Single-shot simulations of dynamic quantum many-body systems
Nat. Phys. {\bf 12}, 451 (2016). 


\bibitem{Lode} A. U. Lode, and C. Bruder, 
%Fragmented superradiance of a Bose-Einstein condensate in an optical cavity. 
Phys. Rev. Lett. \textbf{118}, 013603 (2017). 


\bibitem{c6} B. Chatterjee, and A. U. Lode, 
%Order parameter and detection for crystallized dipolar bosons in lattices 
{\bf arXiv:1708.07409} (2017). 

\bibitem{zorzetos} G. M. Koutentakis, S. I. Mistakidis, and P. Schmelcher, to be submitted.   

\bibitem{filled_vortex} G. C. Katsimiga, S. I. Mistakidis, G. M. Koutentakis, P. G. Kevrekidis, and P. Schmelcher, 
New J. Phys. \textbf{19}, 123012 (2017).  


\bibitem{Naraschewski} M. Naraschewski, and R. J. Glauber, Phys. Rev. A \textbf{59}, 4595 (1999). 


\bibitem{density_matrix} K. Sakmann, A. I. Streltsov, O. E. Alon, and L. S. Cederbaum, 
%Reduced density matrices and coherence of trapped interacting bosons. 
Phys. Rev. A \textbf{78}, 023615 (2008). 


\bibitem{Tavares} P. E. S. Tavares, A. R. Fritsch, G. D. Telles, M. S. Hussein, F. Impens, R. Kaiser, and V. S. Bagnato, 
%Chaotic nature of turbulent Bose-Einstein condensates. arXiv preprint 
arXiv:\textbf{1606.01589} (2016). 


\bibitem{granulation} M. C. Tsatsos, J. H. V. Nguyen, A. U. J. Lode, G. D. Telles, D. Luo, V. S. Bagnato, and R. G. Hulet, 
%Granulation in an atomic Bose-Einstein condensate. arXiv preprint 
arXiv:\textbf{1707.04055} (2017). 


\bibitem{Endres} M. Endres, M. Cheneau, T. Fukuhara, C. Weitenberg, P. Schau\ss, C. Gross, L. Mazza, M. C. Ba\~nuls, 
L. Pollet, I. Bloch, and S. Kuhr, 
%Single-site-and single-atom-resolved measurement of correlation functions. 
Appl. Phys. B \textbf{113}, 27 (2013). 


\bibitem{Proukakis} R. W. Pattinson, T. P. Billam, S. A. Gardiner, D. J. McCarron, H. W. Cho,
S. L. Cornish, N. G. Parker, and N. P. Proukakis, 
Phys. Rev. A {\bf 87}, 013625 (2013).
%Equilibrium solutions for immiscible two-species Bose-Einstein condensates
%in perturbed harmonic traps
  
\bibitem{wen} L. Wen, W. M. Liu, Y. Cai, J. M. Zhang, and J. Hu, Phys. Rev.
A 85, 043602 (2012).  

\bibitem{markus1} A. Weller, J. P. Ronzheimer, C. Gross, J. Esteve, M. K. Oberthaler, D. J. Frantzeskakis, G. Theocharis, and 
P. G. Kevrekidis, Phys. Rev. Lett. {\bf 101}, 130401 (2008).

\bibitem{markus2} G. Theocharis, A. Weller, J. P. Ronzheimer, C. Gross, M. K. Oberthaler, P. G. Kevrekidis, and D. J. 
Frantzeskakis, Phys. Rev. A {\bf 81}, 063604 (2010).

\bibitem{Bisset} R. N. Bisset, P. G. Kevrekidis, and C. Ticknor, 
%Enhanced quantum spin fluctuations in a binary Bose-Einstein condensate. arXiv preprint 
Phys. Rev. A {\bf 97}, 023602 (2018).

\bibitem{gautam} S. Gautam, and S. K. Adhikari,
Phys. Rev. A {\bf 90}, 043619 (2014).
%Phase separation in a spin-orbit coupled Bose-Einstein condensate


  
\end{thebibliography}
\end{document}